\documentclass[twocolumn,pre,superscriptaddress]{revtex4}

\usepackage{graphicx}
\usepackage{amsmath,amssymb}
\usepackage{subfigure}

\newcommand\defn{\textit}

\newcommand\e{\mathrm{e}}
\renewcommand\d{\mathrm{d}}

\newcommand\av[1]{\langle#1\rangle}
\newcommand\etal{\textit{et~al.}}

\newcommand\half{\mbox{$\frac12$}}
\newcommand{\Ord}{\mathrm{O}}

\begin{document}

\title{Competing epidemics on complex networks}

\author{Brian Karrer}
\affiliation{Department of Physics, University of Michigan, Ann Arbor,
MI 48109, U.S.A.}
\author{M. E. J. Newman}
\affiliation{Department of Physics, University of Michigan, Ann Arbor,
MI 48109, U.S.A.}
\affiliation{Center for the Study of Complex Systems, University of
  Michigan, Ann Arbor, MI 48109, U.S.A.}

\begin{abstract}
  Human diseases spread over networks of contacts between individuals and a
  substantial body of recent research has focused on the dynamics of the
  spreading process.  Here we examine a model of two competing diseases
  spreading over the same network at the same time, where infection with
  either disease gives an individual subsequent immunity to both.  Using a
  combination of analytic and numerical methods, we derive the phase
  diagram of the system and estimates of the expected final numbers of
  individuals infected with each disease.  The system shows an unusual
  dynamical transition between dominance of one disease and dominance of
  the other as a function of their relative rates of growth.  Close to this
  transition the final outcomes show strong dependence on stochastic
  fluctuations in the early stages of growth, dependence that decreases
  with increasing network size, but does so sufficiently slowly as still to
  be easily visible in systems with millions or billions of individuals.
  In most regions of the phase diagram we find that one disease eventually
  dominates while the other reaches only a vanishing fraction of the
  network, but the system also displays a significant coexistence regime in
  which both diseases reach epidemic proportions and infect an extensive
  fraction of the network.
\end{abstract}


\maketitle

\section{Introduction}
\label{sec:intro}
Diseases spread over networks of contacts between individuals, and a full
understanding of their nature and behavior requires an understanding of the
relevant network structure and the effects that structure has on the
spreading process.  Traditional theories of disease propagation largely
ignore network effects~\cite{AM91,Hethcote00}, but there has been a
significant volume of research in recent years aimed at building an
understanding of the role that networks
play~\cite{Keeling01,Liljeros01,PV01a,Newman02c,Bansal20071,Kenah2011}.
Network epidemiology ideas have also been applied to cultural transmission
processes, such as the spread of ideas, rumors, fashions, or opinions,
which may occur by mechanisms qualitatively similar in some respects to the
spread of biological disease~\cite{DW04}.

One of the most fundamental and important theoretical models of disease
spreading is the susceptible-infected-recovered (SIR) model, in which
initially susceptible individuals catch the disease of interest from
infected individuals, becoming infected themselves and possibly passing the
disease to others, before recovering and acquiring permanent immunity so
that each individual catches the disease at most once.  As well as being a
reasonably accurate, if simplified, representation of the dynamics of many
real-world diseases, the SIR model is important to network epidemiology
from a theoretical viewpoint because it is closely related to bond
percolation on a given contact
network~\cite{Mollison77,Grassberger83,Sander02,Newman02c,Kenah2011}.  The
mapping is a straightforward one: edges in the network are occupied with
independent probability~$T$, equal to the time-integrated probability that
a neighbor of an infected individual will become infected.  The value
of~$T$ is a property both of the disease and of individuals' behavior, and
is called the \defn{transmissibility} or \defn{infectivity}.  The occupied
edges of the network form connected percolation clusters and the members of
a cluster represent the set of vertices that will become infected in the
limit of long time if any vertex in the cluster is initially infected.  In
general the clusters are small for low values of the transmissibility,
consisting of only few vertices each, but as the transmissibility increases
they become larger and eventually an extensive \defn{giant cluster} forms,
corresponding to a potential epidemic outbreak of the disease in which an
non-negligible fraction of the population becomes infected.  The point at
which the giant cluster forms is a continuous phase transition and is
referred to in the epidemiology literature as the \defn{epidemic
  threshold}.

In this paper we study the behavior of two SIR-type diseases competing for
the same population of hosts and spreading over the same contact network.
In some cases a pair of diseases---or two strains of the same disease, such
as two strains of influenza---spread through the same population and
interact through \defn{cross-immunity}: infection with and recovery from
either strain imparts subsequent immunity to both so that each individual
can catch at most one of the two strains.  The question we address is
whether and how far the two strains will spread under such circumstances,
as a function of parameters such as transmissibility and network structure.

A simple case of this problem was studied previously by
Newman~\cite{Newman05c} using the mapping to bond percolation described
above.  In that work it was assumed that the spread of one disease begins
only after the other has entirely run its course.  The first disease can
then be regarded as effectively removing from the population all
individuals it infects, leaving a pared-down remnant of the original
network, referred to as the \defn{residual network}, for the second disease
to spread on.  If the first disease spreads too readily and consumes too
large a portion of the network, then the residual network will be too sparse
to support the spread of the second disease.  Thus there is an upper limit
on the transmissibility of the first disease if the second is to spread,
which was dubbed the \defn{coexistence threshold} in Ref.~\cite{Newman05c}.

In this paper, we study the more general---and more realistic---case in
which the two diseases spread concurrently.  Although this appears to be a
harder problem it turns out, as we will show, that many of the results of
the earlier analysis can be applied or adapted to the more general
situation.  (A different generalization to the case in which one disease
confers only partial immunity to the other, has been studied by Funk and
Jansen~\cite{Funk2010}.  The case of full cross-immunity for concurrent
diseases following a susceptible-infected-susceptible (SIS) dynamics has
also been studied~\cite{Ahn2006}, but the results are not directly
applicable to the case we study because the SIS model does not map to bond
percolation.)

When studying diseases spreading at well separated times, the precise
dynamics of the diseases is not important for predicting the final outbreak
sizes.  The mapping to percolation tells us everything we need to know and
details like whether one disease spreads faster than the other have no
effect.  When considering the concurrent spread of diseases, however, such
details are important and we must specify what dynamics we are considering.
In this paper we adopt one of the simplest and best-known of dynamical
frameworks for SIR diseases, the Reed--Frost model.  This model uses a
discrete time-step for its evolution and on each step susceptible
individuals have a fixed probability, equal to the transmissibility~$T$, of
being infected by each of their infected network neighbors, so that a
susceptible individual with $k$ infected neighbors remains susceptible with
probability~$(1-T)^k$.  Infected individuals remain infected for exactly
one time-step and then recover, becoming immune and remaining in the immune
state indefinitely thereafter.  Thus the model has just two parameters, the
length of the time-step and the transmissibility.  (The other common choice
for SIR dynamics, the Kermack--McKendrick model, uses continuous time and
stochastically constant rates of infection and recovery, but again has just
two parameters, equal to the two rates.  The Kermack--McKendrick model can
also be mapped to a bond percolation process, but on a directed, rather
than undirected, network, which makes its analysis less transparent, and it
is in part for this reason that we use the Reed--Frost model in this
paper.)

In our studies we make use of a two-disease generalization of the
Reed--Frost model as follows.  The two diseases, which we label red and
blue, each start from a single, randomly chosen, infected vertex.  In the
simplest case the two diseases are assumed to start at the same instant,
although we will relax this condition later.  The diseases have
transmissibilities $T_r$ and~$T_b$, for red and blue respectively.  We also
allow their time-steps to be different, so that one spreads on a faster
clock than the other.  Only the ratio of the time-steps is important for
our results, not the overall time-scale, so we fix the time-step for the
blue disease to be~1 and vary the time-step for the red disease, which we
denote~$\alpha$.  Without loss of generality we assume that
$0\le\alpha\le1$, meaning that red always spreads faster than (or at the
same speed as) blue.

Thus there are three parameters in our two-disease Reed--Frost model,
$T_r$, $T_b$, and~$\alpha$, as opposed to two in the one-disease version.
Despite its simplicity, we will see that this model displays some complex
and interesting behaviors.

\section{Principal results}
\label{sec:summary}
The behavior of the system we study is, as we have said, quite complex, so
we begin in this section with a nontechnical summary.  Detailed arguments
and numerical results are given in the following sections.

Initially we consider the results in the limit of large network size, for
which the behavior of the system is relatively simple.  As we will show,
however, networks must, in some parameter ranges, be very large indeed to
be considered ``large'' in this sense, and real contact networks---even
networks of billions of people---are not large enough.  Thus finite-size
effects can be important under real-world conditions, and so we give an
analysis of these also.

Consider then the behavior of the system on a contact network of $n$
vertices, in the limit of large~$n$.  In order to take the limit we need to
specify how our contact network is defined for given~$n$.  The analytic and
numerical results in this paper are all calculated using the so-called
configuration model~\cite{MR95,NSW01}, i.e.,~a random graph with a
specified degree distribution, and the limit of large network size is taken
as the limit of this model with the degree distribution held
fixed~\cite{note1}.  Qualitatively, the results we report should apply to
other networks as well, subject to some relatively mild conditions, but our
quantitative results are all for the configuration model.

For given values of the three parameters $T_r$, $T_b$, and~$\alpha$, we
consider the fate of our two diseases as they each spread from a single
randomly chosen initial disease carrier.  Let us suppose that the epidemic
threshold for our network---the position of the percolation transition, as
discussed in the introduction---falls at a critical bond
probability~$\phi_c$.  For instance, on configuration model networks it is
known that
\begin{equation}
\phi_c = {\av{k}\over\av{k^2}-\av{k}},
\label{eq:epidthreshold}
\end{equation}
where $\av{k}$ and $\av{k^2}$ represent the mean and mean-square degrees
respectively~\cite{CEBH00,CNSW00}.  If the transmissibility of either of
our diseases falls below this value the disease will not spread, reaching
only $\Ord(1)$ vertices before dying out, in which case that disease can be
ignored and the outcome for the other disease is given by standard
single-disease results.  For nontrivial results, therefore, we require
$T_r,T_b>\phi_c$.

Even for transmissibility above the epidemic threshold it is not guaranteed
that a disease will spread.  A disease starting from a single initial
carrier can still die out, either because it starts in a small component of
the network (not the giant component) or just because of stochastic
fluctuations in the spreading process.  Again, however, this gives trivial
behavior, and so we limit our discussion to cases in which both diseases
``take off,'' meaning that both would, in the absence of the other disease,
ultimately reach epidemic proportions, infecting an extensive fraction of
the network in the limit of large size.

With these assumptions, both diseases spread exponentially at first: each
is surrounded by a sea of susceptible individuals to infect---a naive
population in the epidemiology jargon---and on an arbitrarily large network
the two diseases start arbitrarily far apart and hence do not initially
interfere with one another.  The rate of exponential growth for the blue
disease depends on~$T_b$ and for the red disease on $T_r$ and~$\alpha$.  We
will shortly calculate what these rates are, but for the moment suppose
that we know the rates and that the ratio of the rate for red to the rate
for blue is~$\beta$.  Then on average the number of individuals infected by
blue goes as $N_b = \e^{rt}$ for some rate~$r$ (with the overall multiplier
fixed by the condition that $N_b=1$ at $t=0$) and the number infected by
red goes as $N_r = \e^{\beta rt} = N_b^\beta$.

Now suppose that $\beta<1$, meaning that red grows slower and blue grows
faster, and let us wait a certain amount of time until blue grows to
fill~$\Ord(n)$ vertices, meaning that $N_b=cn$ for some small constant~$c$.
In the same amount of time red infects $N_r = N_b^\beta = (cn)^\beta$
vertices, which is a fraction $(cn)^\beta/n = c^\beta/n^{1-\beta}$ of the
entire network, i.e.,~a vanishing fraction in the limit of large~$n$ since
$\beta<1$.  Thus, as far as the blue disease is concerned, red is
irrelevant---cross-immunity of individuals previously infected with the red
disease is a negligible impediment to blue's spread, and blue's dynamics
will be the same as if it were the only disease.  We can make a similar
argument for the case where $\beta>1$ and red grows faster, showing that in
the time red takes to reach~$\Ord(n)$ vertices blue reaches
only~$\Ord(n^{1/\beta})$, or a fraction $\Ord(1/n^{1-1/\beta})$, which is
again negligible in the limit of large~$n$.

This insight is important because it allows us to treat our concurrent
diseases as if they were in fact spreading non-concurrently, one after the
other.  The faster disease spreads to the entire network, infecting
essentially everyone it was going to infect, before the slower disease
rises beyond the level of insignificance.  This means that, in the limit of
large~$n$, the eventual outcome for the two diseases can be predicted using
static percolation arguments of the type employed for time-separated
diseases in the work of Newman~\cite{Newman05c} discussed in the
introduction.

The value~$\beta=1$ at which the growth rates of the two diseases are equal
forms a surface in the three-dimensional parameter space~$(T_r,T_b,\alpha)$
of the model.  On one side of this surface the blue disease grows faster
and is equivalent to the first disease in the time-separated picture of
Ref.~\cite{Newman05c}, while the red disease plays the role of the second
disease.  On the other side of the surface the roles are reversed.

\begin{figure}
\begin{center}
\includegraphics[width=\columnwidth]{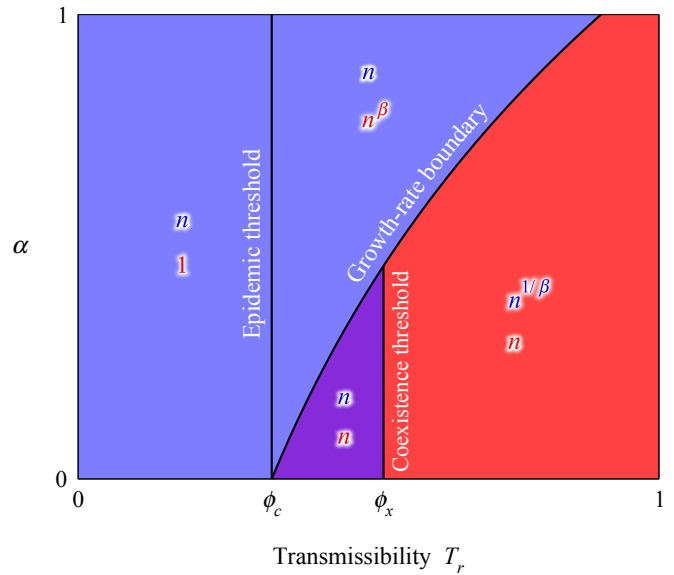}
\end{center}
\caption{A two-dimensional slice through the phase diagram of the system,
  with $T_b$ held constant while $T_r$ and~$\alpha$ vary.  The area colors
  represent the dominant disease in each phase, while the colored symbols
  give the leading-order scaling of the expected number of individuals
  infected with each disease.}
\label{fig:phase}
\end{figure}

Figure~\ref{fig:phase} shows a two-dimensional slice of the parameter space
for fixed~$T_b$ and varying values of $T_r$ and~$\alpha$.  The surface
$\beta=1$ is represented by the diagonal curve, colored blue on one side
and red on the other to represent the dominant disease.  If we are in the
regime $\beta>1$, which is to the right of the curve, then red will always
produce an epidemic.  When red has run its course and finished spreading
(but blue has still reached only a vanishing fraction of the network), it
leaves behind it a residual network, as described earlier, and blue will
spread further only if that residual network is sufficiently dense that the
threshold for bond percolation on it falls below~$T_b$.  Just as before,
this places an upper limit on the value of~$T_r$ if both diseases are to
spread, which we call the coexistence threshold and denote~$\phi_x$.  If
$T_r<\phi_x$ then blue can spread and cause its own epidemic; if not, then
blue only ever infects $\Ord(n^{1/\beta})$ vertices.  The
value~$T_r=\phi_x$ constitutes another surface in the parameter space,
which appears as a vertical straight line in Fig.~\ref{fig:phase},
separating the coexistence regime (colored purple) from the regime where
there is only one epidemic (red).  Note that the value of~$\phi_x$ depends
on the transmissibility~$T_b$ of the blue disease---a more virulent blue
disease may spread on a residual network were a less virulent one would
fail---so the position of the vertical line will shift as we take different
slices through the parameter space~\cite{note2}.

One might imagine that one would see a similar behavior on the other
(left-hand) side of the curve in Fig.~\ref{fig:phase}.  On this side the
blue disease spreads faster and will always cause an epidemic, but if its
transmissibility is sufficiently low, one might argue, then it could leave
behind a residual network dense enough to allow the red disease to
subsequently spread and also cause an epidemic.  It turns out, however,
that this does not occur.  As we will show, a necessary condition for
coexistence of the slower red disease in this case is that $T_r>T_b$.
However, given that~$\alpha\le1$, it turns out that we also require
$T_b>T_r$ if blue is to be the faster disease in the first place, and hence
we have a contradiction and coexistence is not possible.  Thus there is no
coexistence region to the left of the curve.

Figure~\ref{fig:phase} is completed by a third line, again vertical,
representing the point at which $T_r$ falls below the percolation
threshold~$\phi_c$.  To the left of this line, the red disease does not
cause an epidemic under any circumstances, infecting only~$\Ord(1)$
individuals.

The complete figure constitutes a phase diagram for this system and has
four regions, in one of which we have coexistence of the two diseases,
while in each of the others one disease dominates and the other disease
reaches only a vanishing fraction of the network.  Note that the
percolation and coexistence thresholds, which were identified
previously~\cite{Newman05c}, correspond to continuous phase transitions in
the sizes of the epidemics, but the curved $\beta=1$ boundary marks a
different kind of transition at which the sizes jump discontinuously.  This
transition, which is a purely dynamical phenomenon driven by the differing
exponential growth rates of the diseases, was not present in previous
studies.

An interesting feature of the phase diagram is that there can be a maximum
value of $\alpha$ above which coexistence does not occur---note that the
purple coexistence region extends only up to a certain value of~$\alpha$
and not beyond.  Thus, for instance, for a random graph with a Poisson
degree distribution, it turns out that coexistence requires
$\alpha\le\frac12$, meaning that the red disease must spread at least twice
as fast as the blue disease.

While still treating the large-size limit, we can consider other variants
of our model.  In particular, the assumption that both diseases start
spreading at the same moment is unrealistic and can be relaxed.  In most
real-world cases one disease (or one strain of disease) starts spreading
before the other, which is equivalent to having both start at $t=0$ but
having one disease start with more than one carrier while the other starts
with just one.  As long as the number of initial carriers for both diseases
is constant, however, this makes no difference in the limit of large
network size because the disease with the larger exponential growth rate
will always win in the end.  Asymptotically, therefore, the phase diagram
and other properties are unchanged.

The picture changes, however, when we consider a network of finite size.
For diseases that start at the same moment we can make the same argument as
before: in the time it takes for the faster-growing disease to spread
to~$\Ord(n)$ vertices the slower-growing one spreads to~$\Ord(n^{\beta})$
or $\Ord(n^{1/\beta})$, as appropriate.  Close to the $\beta=1$ boundary,
however, $\Ord(n^\beta)$~could be a large number, so there will be some
rounding of the previously discontinuous transition as we cross the
boundary.  (In this sense the transition is unlike a traditional
first-order phase transition, which remains discontinuous in a finite-sized
system but shows fluctuations in its position.  In the present model the
position of the transition is unchanged but it is no longer discontinuous.)
The other transitions, which are continuous phase transitions, will also
show rounding, as is typical of such transitions in finite systems.

A more dramatic difference appears if, as above, we allow the diseases to
start their spread at different times, or equivalently give one disease a
larger number of initial carriers than the other.  In particular, if we
give a head start to the disease with the slower exponential growth rate
then, depending on the size of that head start and the difference in rates,
it is possible that the slower disease will, on the finite network, reach a
significant fraction of the vertices in advance of the faster disease.  The
faster disease simply does not have enough time to overtake the slower
one's lead and exclude it from the network.  In this case it is possible
for the slower disease to actually exclude the faster one, imparting
immunity upon a sufficiently large fraction of individuals as to prevent an
epidemic of the faster disease, a complete reversal of the behavior in the
infinite-size limit.

Suppose for example that the blue disease is faster, meaning that
$\beta<1$.  As before the blue disease starts with a single carrier at
$t=0$ and the number grows as $N_b = \e^{rt}$ on average for some
constant~$r$.  But let us now suppose that red has $A>1$ infected carriers
at $t=0$, so that its number of cases is not $\e^{\beta rt}$ but $N_r =
A\e^{\beta rt} = AN_b^\beta$.  The faster blue disease will catch up to the
slower red one at the point where $N_b=N_r$ or equivalently $N_b =
AN_b^\beta$, which can be rearranged to give $N_b = A^{1/(1-\beta)}$.  If
blue is to catch up to red, this point must fall well before we reach the
size~$n$ of the entire network and the exponential growth of the diseases
is exhausted.  In other words, blue catches up only if
\begin{equation}
n \gg A^{1/(1-\beta)}.
\label{eq:howlarge}
\end{equation}
We can make a similar argument if $\beta>1$ and red is the faster disease.
The result is the same but with $\beta$ replaced by~$1/\beta$.

But note that if the growth rates of the two diseases are similar, so that
$\beta$ is close to~1, then the exponent in Eq.~\eqref{eq:howlarge} becomes
large, meaning that the network needs to be very large to prevent the
slower disease from dominating.  With $\beta=0.9$, for instance, and a
modest value of $A=10$, we would need $n\gg10^{10}$, which is several
orders of magnitude larger than the population of the Earth.  Thus the
finite-size effects can be very much noticeable for populations of
realistic sizes.

Moreover, we have throughout this argument ignored stochastic fluctuations
in the growth process.  In the early stages of development of an epidemic,
when only a handful of individuals are infected, fluctuations can be large
as a fraction of epidemic size---easily a factor of two or more---and this
can either increase or decrease the lead that one disease has over another,
or create a lead where none existed previously.  Even if our two diseases
start spreading at exactly the same moment, a lead gained by the slower
disease, by virtue of chance fluctuations early in the process, can be
enough, particularly when $\beta$ is close to~1, to allow it to dominate,
or at least coexist with, the other disease, in defiance of the predictions
of the infinite-$n$ theory.  This behavior means that the outcome is
broadly undetermined for random initial conditions on finite-sized networks
within certain regions of the parameter space.  (We note that this is quite
different from the behavior of a single disease modeled with the SIR model,
for which initial fluctuations have no effect whatsoever on the final
number of individuals infected, assuming that the disease does not die
out.)

On the other hand, as we will see, if we wait some time until one disease
has grown to fill a small fraction of the network and the period of gross
initial fluctuations has ended, then it is possible to predict the final
outcome quite accurately from arguments such as those above.  One would not
be far off the mark if one where to assume the dynamics to be governed by a
deterministic growth process after waiting a suitable initial period for
the fluctuations to become small.  In this regard, our results agree with
recent work by Marceau~\etal~\cite{Marceau2011}, who consider a
deterministic differential equation model of competing diseases and show
that it is able to predict final outcomes with some accuracy, but only
given an initial condition that specifies the fraction of individuals
infected with each disease at a time after fluctuations have become
negligible.  Thus a deterministic approximation does seem to work in this
regime, although it also misses some of the most interesting phenomena
exhibited by the system.

In the following sections we discuss the results above in more detail,
giving a combination of analytic derivations and numerical results to
motivate our conclusions.

\section{Epidemic growth rates}
\label{sec:exponential}
Our first step in demonstrating the results summarized in
Section~\ref{sec:summary} is to derive the exponential growth of each of
the two diseases in the absence of the other.  In the epidemiology
literature the growth of diseases is usually parametrized by the
\defn{basic reproductive ratio}~$R_0$, which is the average number of
additional infections caused by a newly infected individual.  Consider a
vertex in our network model that is infected by a disease with
transmissibility~$T$ and suppose the vertex has $k$ network neighbors
excluding the one that gave it the disease in the first place (which is
obviously already infected)---$k$~is the so-called \defn{excess degree}.
At early times the neighbors will, with high probability, all be
susceptible and each becomes infected with independent probability~$T$, so
that the expected number of infectees is~$Tk$.  The value of~$R_0$ is the
average of this quantity over the excess degree~$k$.

Let $p_k$ be the degree distribution of the network, i.e.,~$p_k$~is the
fraction of vertices with degree~$k$.  The excess degree, however, is
distributed not according to this distribution, but according to its own
\defn{excess degree distribution}~\cite{NSW01}
\begin{equation}
q_k = {(k+1)p_{k+1}\over\av{k}},
\label{eq:excess}
\end{equation}
where $\av{k}=\sum_k kp_k$ is, as before, the mean degree in the network.
Performing the average over the excess degree, the basic reproductive ratio
is now given by
\begin{align}
R_0 &= T\sum_{k=0}^\infty kq_k
     = {T\over\av{k}} \sum_{k=0}^\infty k(k-1)p_k \nonumber\\
    &= T {\av{k^2}-\av{k}\over\av{k}}.
\label{eq:brr}
\end{align}
The epidemic threshold, which separates the regime in which the disease
dies out from the regime in which it grows exponentially, is the point at
which $R_0=1$, or equivalently the point at which $T = \phi_c =
\av{k}/(\av{k^2}-\av{k})$, as in Eq.~\eqref{eq:epidthreshold}.

In the Reed--Frost model in a naive population, each infected individual
infects, on average, $R_0$~susceptibles on each time-step, then recovers
and is no longer infectious.  Thus $R_0$ is precisely the average factor by
which the number of infected individuals changes on each step.  If we
define
\begin{equation}
R_b = T_b {\av{k^2}-\av{k}\over\av{k}} = {T_b\over\phi_c}
\end{equation}
to be the reproductive number for the blue disease in our two-disease
system, and recall that the blue disease by hypothesis executes exactly one
time-step per unit time, then in a naive population the average number of
individuals infected with the blue disease at early times~$t$ is
\begin{equation}
N_b = R_b^t = \e^{t\ln R_b},
\label{eq:nbtime}
\end{equation}
where the outbreak is assumed to start from a single infected carrier
at~$t=0$~\cite{note3}.

Similarly for the red disease we can define
\begin{equation}
R_r = T_r {\av{k^2}-\av{k}\over\av{k}} = {T_r\over\phi_c}.
\end{equation}
But recall that the red disease executes one time-step every $\alpha$ units
of time, which means that the average number of individuals infected with
red at early times is
\begin{equation}
N_r = R_r^{t/\alpha} = \e^{(t/\alpha)\ln R_r}.
\label{eq:nrtime}
\end{equation}
Thus both diseases display exponential growth but with different growth
rates: $\ln R_b$ for blue and $\alpha^{-1} \ln R_r$ for red.  Note that
these rates are both positive provided both diseases are above the epidemic
threshold, so that $R_b,R_r>1$.

We define $\beta$ as before to be the ratio of the exponential growth rates
of the two diseases:
\begin{equation}
\beta = \frac{\ln R_r}{\alpha \ln R_b}
      = \frac{\ln (T_r/\phi_c)}{\alpha \ln (T_b/\phi_c)},
\label{eq:defsbeta}
\end{equation}
which we can calculate explicitly given the values of $T_b$, $T_r$,
$\alpha$, and the degree distribution of the network.  Now, by the argument
given in Section~\ref{sec:summary}, we can show that for a network of $n$
vertices with $n$ large, the faster-growing disease will fill a
fraction~$\Ord(n)$ of the network before the slower one has a chance to
fill more than a vanishing fraction.  Therefore the faster disease can be
treated as if the slower one did not exist and the analysis for
non-concurrent diseases, spreading one after another, applies.  (As we
pointed out in Section~\ref{sec:summary}, other behaviors are possible on
finite networks, but for the moment let us focus on the large-$n$ limit.)

Equation~\eqref{eq:defsbeta} allows us to calculate the position of the
diagonal curve in Fig.~\ref{fig:phase}, which we call the \defn{growth-rate
  boundary}, separating the region where the blue disease dominates from
the region where the red one does.  This boundary corresponds, as we have
seen, to $\beta=1$, and hence is given by
\begin{equation}
\alpha = {\ln(T_r/\phi_c)\over\ln(T_b/\phi_c)}.
\label{eq:grboundary}
\end{equation}
Note that this implies that the boundary meets the $\alpha=0$ axis at the
epidemic threshold $T_r=\phi_c$ for the red disease, as depicted in
Fig.~\ref{fig:phase}.

\section{Percolation analysis}
\label{sec:bondp}
Given that the two diseases can be treated as being well separated in time,
we can now apply a bond percolation analysis to derive a number of useful
results concerning the transmissibilities and the positions of the various
transitions depicted in the phase diagram of Fig.~\ref{fig:phase}.  Such an
analysis was described previously in Ref.~\cite{Newman05c}, whose results
we review and extend in this section.

Consider our two diseases spreading on a configuration model network with
degree distribution~$p_k$ and $n$ vertices, where $n$ is again large, and
let us suppose, for the sake of argument, that the red disease has the
faster rate of growth, so that it can be treated as running its course
first, to be followed later by the blue disease.  (The case where blue
grows faster can be treated by similar arguments, but with the colors
reversed.)  We thus consider a bond percolation process on our network with
bond occupation probability equal to the transmissibility~$T_r$ of the red
disease.  The size of an epidemic outbreak of the disease, the number of
people affected by the disease in the limit of long time, is then equal to
the size of the giant percolation cluster, as described in the
introduction.

Consider any vertex in the network and let $u$ be the probability that it
is not connected to the giant cluster by a particular one of its edges.
This can happen in either of two ways.  First, the edge could be
unoccupied, which occurs with probability~$1-T_r$.  Second, the edge could
be occupied (probability~$T_r$) but the vertex at its other end does not
belong to the giant cluster.  The latter happens if and only if none of
that vertex's other edges connect it to the giant cluster, which occurs
with probability~$u^k$, where $k$ is the number of other edges, the
quantity we call the excess degree.  Thus the total probability is
$1-T_r+T_ru^k$.  Now we average this expression over the distribution~$q_k$
of the excess degree, Eq.~\eqref{eq:excess}, to get a complete expression
for~$u$ thus:
\begin{equation}
u = \sum_{k=0}^\infty q_k (1 - T_r + T_r u^k)
  = 1 - T_r + T_r F_1(u),
\label{eq:percu}
\end{equation}
where $F_1(z) = \sum_k q_k z^k$ is the probability generating function for
the excess degree distribution and we have made use of the normalization
condition $\sum_k q_k = 1$.

If we can solve Eq.~\eqref{eq:percu} for~$u$, we can use the result to
calculate the size of the giant cluster.  A randomly chosen vertex of
(total) degree~$k$ is not in the giant cluster if none of its edges connect
it to that cluster, which happens with probability~$u^k$ again.  Now,
however, $k$~is distributed according to the total degree
distribution~$p_k$, so the average probability of being outside the giant
cluster is $\sum_k p_k u^k = F_0(u)$ and the probability of being in the
giant cluster is
\begin{equation}
S = 1 - F_0(u),
\label{eq:percs}
\end{equation}
where $F_0(z) = \sum_k p_k z^k$ is the generating function for the degree
distribution.  But this probability is also equal to the expected fraction
of the network occupied by the giant cluster, and hence $S$ gives us the
size of the giant cluster as a fraction of~$n$.

The portion of the network not in the giant cluster, which may include more
than one network component, is the portion we previously called the
residual network.  It is the portion not infected by the red disease and
hence not immune to subsequent infection by the blue disease.  To find out
whether the blue disease will cause a second epidemic, and if so how many
individuals it will infect, we must now perform a second bond percolation
calculation on this residual network.

It is straightforward to show that the residual network itself constitutes
another configuration model network, i.e.,~a random graph with a given
degree distribution, but the degree distribution is different in general
from that of the original network because in removing the vertices infected
by the red disease we reduce the degrees of their uninfected neighbors.
The degree distribution of the residual network cannot be expressed simply
in closed form but, as shown in~\cite{Newman05c}, its generating function,
which we denote~$G_0(z)$, can:
\begin{equation}
G_0(z) = \frac{F_0\bigl(u+(z-1)F_1(u)\bigr)}{F_0(u)},
\label{eq:defsg0}
\end{equation}
where $F_0$, $F_1$, and~$u$ are defined as previously.  Similarly, the
generating function for the excess degree distribution on the residual
network is
\begin{equation}
G_1(z) = \frac{F_1\bigl(u+(z-1)F_1(u)\bigr)}{F_1(u)}.
\label{eq:defsg1}
\end{equation}

Given these expressions the percolation properties of the residual network
are straightforward to calculate.  Consider a bond percolation process on a
configuration model having the degree distribution implied by
Eqs.~\eqref{eq:defsg0} and~\eqref{eq:defsg1}, with bond occupation
probability~$T_b$.  We consider any vertex in the network and let $v$ be
the probability that it is not connected to the giant cluster of the new
percolation process by a particular one of its edges.  Then, by the same
argument as before, $v$~is a solution of
\begin{equation}
v = 1 - T_b + T_b G_1(v),
\end{equation}
and the fraction~$C$ of the residual network filled by the giant cluster is
\begin{equation}
C = 1 - G_0(v).
\end{equation}
Since the residual network is itself a fraction~$1-S$ of the original
network, the giant cluster of the second percolation process fills a
fraction $C(1-S)$ of the original~$n$ vertices.

We can also calculate the positions of the epidemic thresholds for the two
diseases.  They are given by Eq.~\eqref{eq:epidthreshold} with the degree
averages calculated for the entire network or the residual network, as
appropriate.  Alternatively, the thresholds can be expressed in terms of
the generating functions.  The basic reproductive ratio is given by
Eq.~\eqref{eq:brr} to be
\begin{equation}
R_r = T_r \sum_{k=0}^\infty kq_k = T_r F_1'(1),
\end{equation}
where the prime indicates a derivative.  The epidemic threshold corresponds
to the point $R_r=1$, i.e.,~the point where $T_r$ is equal to $1/F_1'(1)$.
(It is straightforward to show that this gives the same result as
Eq.~\eqref{eq:epidthreshold}.)  Similarly the threshold for the blue
disease on the residual network is
\begin{equation}
\phi_b = {1\over G_1'(1)} = {1\over F_1'(u)},
\label{eq:coexist}
\end{equation}
where we have used Eq.~\eqref{eq:defsg1}.  Note that the value of $u$
depends on~$T_r$ via Eq.~\eqref{eq:percu} and hence so does the value
of~$\phi_b$.  Thus Eq.~\eqref{eq:coexist} also implicitly defines the
coexistence threshold~$\phi_x$ as the value of~$T_r$ at which $\phi_b=T_b$
and the blue disease fails to spread.  That is, the coexistence threshold
is the value~$\phi_x$ such that $T_b=1/F_1'(u)$ when $u$ is the solution of
$u = 1 - \phi_x + \phi_x F_1(u)$.  Rearranging both expressions, we could
alternatively write
\begin{equation}
\phi_x = {1-u\over1-F_1(u)},
\label{eq:phix}
\end{equation}
where $u$ satisfies $F_1'(u) = 1/T_b$.

In~\cite{Newman05c} it was shown always to be the case that
$\phi_b>\phi_c$---the threshold for the second disease to spread is greater
than the threshold for the first.  However, a stronger result also holds,
which will be important for our work: it turns out that $\phi_b$ must be
greater also than~$T_r$.  To see this, we first rearrange
Eq.~\eqref{eq:percu} to give an expression for~$T_r$ thus:
\begin{equation}
T_r = {1 - u\over1 - F_1(u)},
\label{eq:tru}
\end{equation}
(which is similar to Eq.~\eqref{eq:phix}, except that $u$ may now take any
value, where in~\eqref{eq:phix} it takes the value that satisfies $F_1'(u)
= 1/T_b$).

Recall that $u$ represents the probability that an edge connects to a
vertex not in the giant cluster of the red disease, and this probability
can only decrease (or stay the same) when~$T_r$ increases, which implies
that $\d T_r/\d u\le0$.  Performing the derivative, we then find that
$F_1(u) - 1 + (1-u)F_1'(u)\le0$, and rearranging this expression we have
\begin{equation}
T_r = {1 - u\over1 - F_1(u)} \le {1\over F_1'(u)} = \phi_b.
\end{equation}
Since $T_b>\phi_b$ for the second disease to spread, this result implies
that $T_b>T_r$ also.  Thus the second disease must have a higher
transmissibility than the first for coexistence to occur.  This means, for
instance, that we cannot have coexistence of two diseases with the same
transmissibility.

We can make similar arguments if the roles of red and blue diseases are
reversed.  If blue is the faster growing then it spreads first and red
spreads subsequently over the residual network blue leaves behind.  The
same equations apply for the sizes of the epidemics and the thresholds, and
we can show that $T_r>T_b$ is a necessary condition for red to spread
extensively.  In this regime, however, the quantity~$\beta$,
Eq.~\eqref{eq:defsbeta}, is by definition less than unity and hence
$\ln(T_r/\phi_c) < \alpha \ln(T_b/\phi_c)$.  Rearranging, we then have
\begin{equation}
\ln T_r < \alpha \ln T_b + (1-\alpha) \ln\phi_c \le \ln T_b,
\end{equation}
and hence $T_r<T_b$.  Thus it is never the case in this regime that
$T_r>T_b$ and so, as claimed in Section~\ref{sec:summary}, coexistence is
not possible when blue is the faster growing disease.

Thus one is left with slightly contrary criteria for coexistence.  On the
one hand we must be in the regime where red is the faster growing disease,
but on the other hand blue must have the higher probability of
transmission.  It is possible this type of behavior may not be common for
real diseases---it seems reasonable that the disease with higher
transmissibility would grow faster, not slower---and hence it may be that
the conditions for coexistence of competing diseases in a single population
are met rather rarely.

As a concrete example of the results of this section, consider a faster red
disease and a slower blue one spreading on a network with the Poisson
degree distribution $p_k = e^{-c}c^{k}/k!$.  In this case $F_0(z) = F_1(z)
= e^{c(z-1)}$---so the degree and excess degree distributions are the
same---and $\phi_c = 1/F_1'(1) = 1/c$.  The coexistence threshold~$\phi_x$
is then given by Eq.~\eqref{eq:phix} where $u$ is the solution of
$F_1'(u)=1/T_b$, or
\begin{equation}
u = 1 - {1\over c} \ln cT_b.
\end{equation}
Substituting this value into~\eqref{eq:phix} gives
\begin{equation}
\phi_x = {T_b\ln cT_b\over cT_b-1}.
\label{eq:poissonphix}
\end{equation}

We can also use the results of this section to find the maximum value
of~$\alpha$ beyond which coexistence does not occur.  Referring to
Fig.~\ref{fig:phase}, we see that the maximum falls on the growth-rate
boundary defined by Eq.~\eqref{eq:grboundary}, at the point where
$T_r=\phi_x$.  For the Poisson degree distribution above, for example,
Eq.~\eqref{eq:grboundary} becomes
\begin{equation}
\alpha = {\ln cT_r\over\ln cT_b},
\end{equation}
and substitution from Eq.~\eqref{eq:poissonphix} then gives
\begin{equation}
\alpha_\textrm{max} = 1 - {\ln(cT_b-1) - \ln\ln cT_b\over\ln cT_b}.
\end{equation}
This is a monotone decreasing function of $T_b$ and has a maximum value
of~$\frac12$ as $T_b\to1/c$ from above.  (The blue disease cannot have a
transmissibility~$T_b$ smaller than $1/c$ and still spread, since $1/c$ is
the percolation threshold for this network.)  Thus, for the Poisson degree
distribution, coexistence requires the red disease to transmit at least
twice as fast as the blue disease---or faster for larger $T_b$ or~$c$.
Similar calculations can be performed for other degree distributions too,
although they require more effort and closed-form expressions are not
always possible.

\section{Numerical results}
\label{sec:simulation}
The results above apply to the case of infinite system size but, as we
argued in Section~\ref{sec:summary}, finite-size effects can be important
in this system even for the largest networks that occur in real-world
situations.  To investigate the behavior of the system on networks of
finite size and compare with the heuristic arguments of
Section~\ref{sec:summary}, we have performed extensive numerical
simulations of the model.  For these simulations we again use random graphs
with Poisson degree distributions and fix the mean degree at $c=3$ in all
cases.  (Small values of $c$ make the coexistence region larger and thus
more easily visible.)  As with our analytic calculations, we start each
disease with a single infected carrier chosen uniformly at random, all
other vertices being initially susceptible.  The diseases start at the same
instant and spread according to the Reed--Frost dynamics defined in
Section~\ref{sec:intro}.  Instances in which one or both of the diseases
die out are discarded~\cite{note4} and we collect statistics for the final
number of individuals infected with each disease for a range of values of
the parameters and for network sizes up to a million vertices.

\begin{figure}
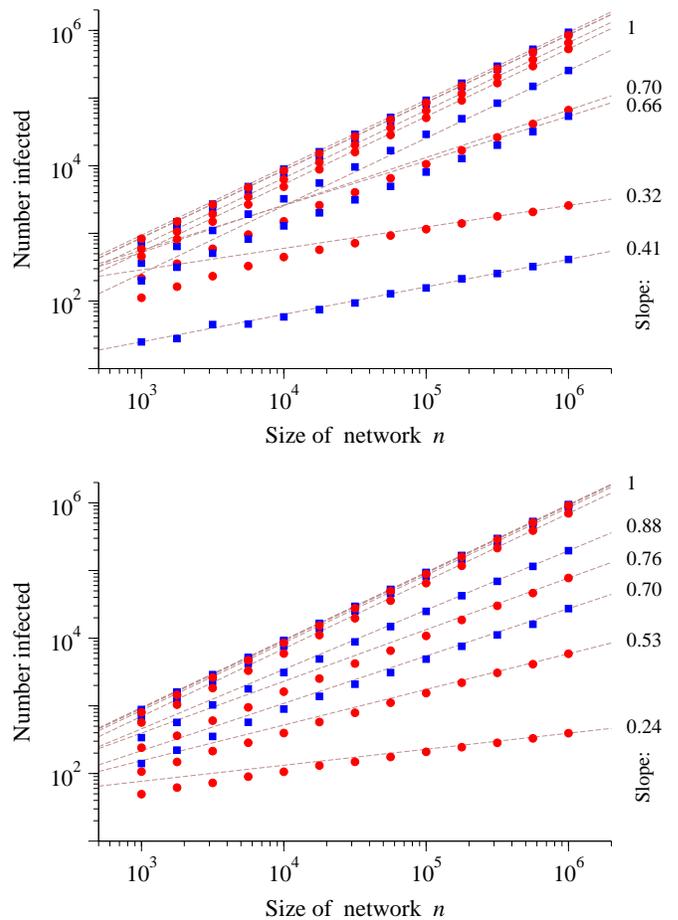

\begin{center}
\includegraphics[width=\columnwidth]{scaling1.eps}\\
{\ }\\
\includegraphics[width=\columnwidth]{scaling2.eps}
\end{center}
\caption{Simulation results for the number of vertices infected with each
  disease as a function of network size for various values of the
  parameters.  Red circles indicate values for the red disease, blue
  squares for the blue disease.  Each point represents an average over 1000
  networks.  In the top panel $\alpha=0.3$, $T_b=1$, and $T_r$ is 0.37,
  0.42, 0.50, 0.55, and 0.75 for the various data sets shown.  These
  parameter values are all outside the coexistence region, in the regime
  where one disease or the other always dominates.  In the lower panel
  $\alpha=0.7$, $T_b=1$, and $T_r$ is 0.4, 0.5, 0.6, 0.8 and~1.  These
  parameters include parts of the coexistence region, where both diseases
  can infect an extensive number of vertices.  The dotted lines represent
  the scaling expected from the arguments given in the text.}
\label{fig:scaling}
\end{figure}

Figure~\ref{fig:scaling} shows the scaling of the numbers of infected
individuals with network size for two different values of the
parameter~$\alpha$ (top and bottom panels in the figure).  The red and blue
points indicate the numerical results for the red and blue diseases,
averaged over a thousand networks each, while the solid lines represent the
expected slope of the scaling in the large-$n$ limit.  As we can see, in
most cases the numerical results appear to converge to the expected slope
as $n$ becomes large, confirming our analytic calculations.  For parameter
values that fall close to the growth-rate boundary, however, the agreement
is poorer (lines of slope 0.70 and 0.66 in the upper panel and 0.76 in the
lower panel).  We interpret this as an effect of the early-time
fluctuations discussed in Section~\ref{sec:summary}.  When we are close to
the boundary, fluctuations in the early growth of one or both diseases can
give the slower-growing disease a lead over the faster one large enough
that the faster one never catches up.  As a result the slower disease
achieves $\Ord(n)$ scaling where normally it would have the
lesser~$\Ord(n^\beta)$ [or~$\Ord(n^{1/\beta})$].  When averaged over many
simulations, this means that the apparent scaling of the number of infected
individuals will lie somewhere between the expected $\Ord(n^\beta)$ and the
steeper~$\Ord(n)$, which is what we see in Fig.~\ref{fig:scaling}.

In this regime, therefore, the observed behavior is a result of initial
fluctuations coupled with finite-size effects.  It should be possible to
reduce the magnitude of the effect either by increasing the system size (so
as to reduce finite-size effects) or by increasing the number of initial
carriers of the disease (to reduce fluctuations).  As we argued in
Section~\ref{sec:summary}, however, one may have to increase the system
size enormously---far beyond our ability to perform the simulations---in
order to make a significant difference.  And reducing early fluctuations by
increasing the number of initial carriers has the undesirable result of
increasing the finite-size effects, since it reduces the length of the
exponential growth phase for both diseases, and therefore again requires us
to increase the system size to get comparable results.  In practice,
therefore, there is no easy way to reach the asymptotic scaling regime when
we are close to the growth-rate boundary.

\begin{figure}
\begin{center}
\includegraphics[width=8cm]{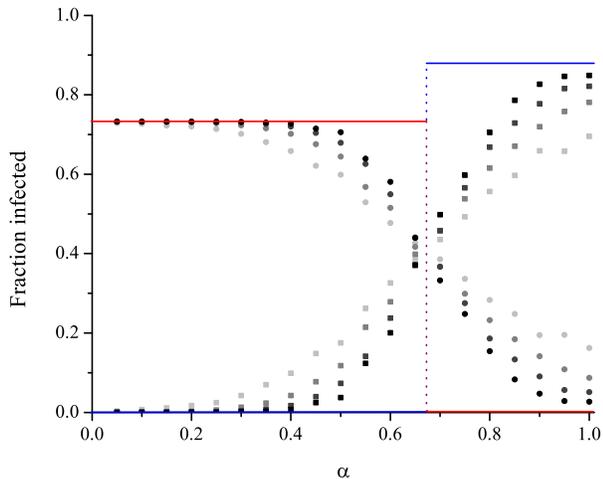}
\end{center}
\caption{Simulation results for the average fraction of vertices infected
  with red (circles) and blue (squares) diseases as a function of~$\alpha$
  for the case $T_r=0.6$, $T_b=0.8$.  Solid lines represent the analytic
  predictions for the same parameter values.  The parameters fall outside
  the coexistence region for all values of~$\alpha$, so one disease or the
  other should always dominate, at least in the limit of large~$n$, and
  there should be a single discontinuous transition as we cross the
  growth-rate boundary.  The shades of the data points indicate network
  sizes of $n=10^3$, $10^4$, $10^5$, and~$10^6$ (lightest points to
  darkest).  Each point is an average over at least $100$ networks.}
\label{fig:abovecoexist}
\end{figure}

In Fig.~\ref{fig:abovecoexist} we show results for the final fraction of
individuals infected with each disease as a function of $\alpha$ for fixed
$T_b$ and~$T_r$.  The value of $T_r$ is chosen to fall above the
coexistence threshold~$\phi_x$, so that we are in the regime on the right
of the phase diagram, Fig.~\ref{fig:phase}, where one disease, either red
or blue, always wins and the other reaches a negligible fraction of
vertices.  The points on the plot represent the numerical results, while
the red and blue lines show our analytic predictions from
Eqs.~\eqref{eq:percu} and~\eqref{eq:percs}.  The lines are horizontal
because the sizes of the epidemics depend only on the transmissibilities
(which are fixed) and not on~$\alpha$, except at the growth-rate boundary,
which is clearly visible as the step where the dominant disease switches
from red to blue.

As we can see from the figure the agreement between analytic and numerical
results is good away from the growth-rate boundary, as we expect.  Away
from the boundary finite-size effects are small and fluctuations should
have a negligible impact on outcomes.  Closer to the boundary agreement is
poorer because, once again, the plotted points represent averages over many
simulations and in some of those simulations the ``wrong'' disease
dominates because of chance fluctuations (or merely fails to be vanishingly
small), moving the average away from the infinite-$n$ prediction.
Moreover, in this regime convergence to the analytic prediction with
increasing~$n$ appears slow (represented by the varying shades of gray
among the points), which accords qualitatively with our expectations from
Eq.~\eqref{eq:howlarge} and the accompanying argument.

\begin{figure}
\begin{center}
\includegraphics[width=8cm]{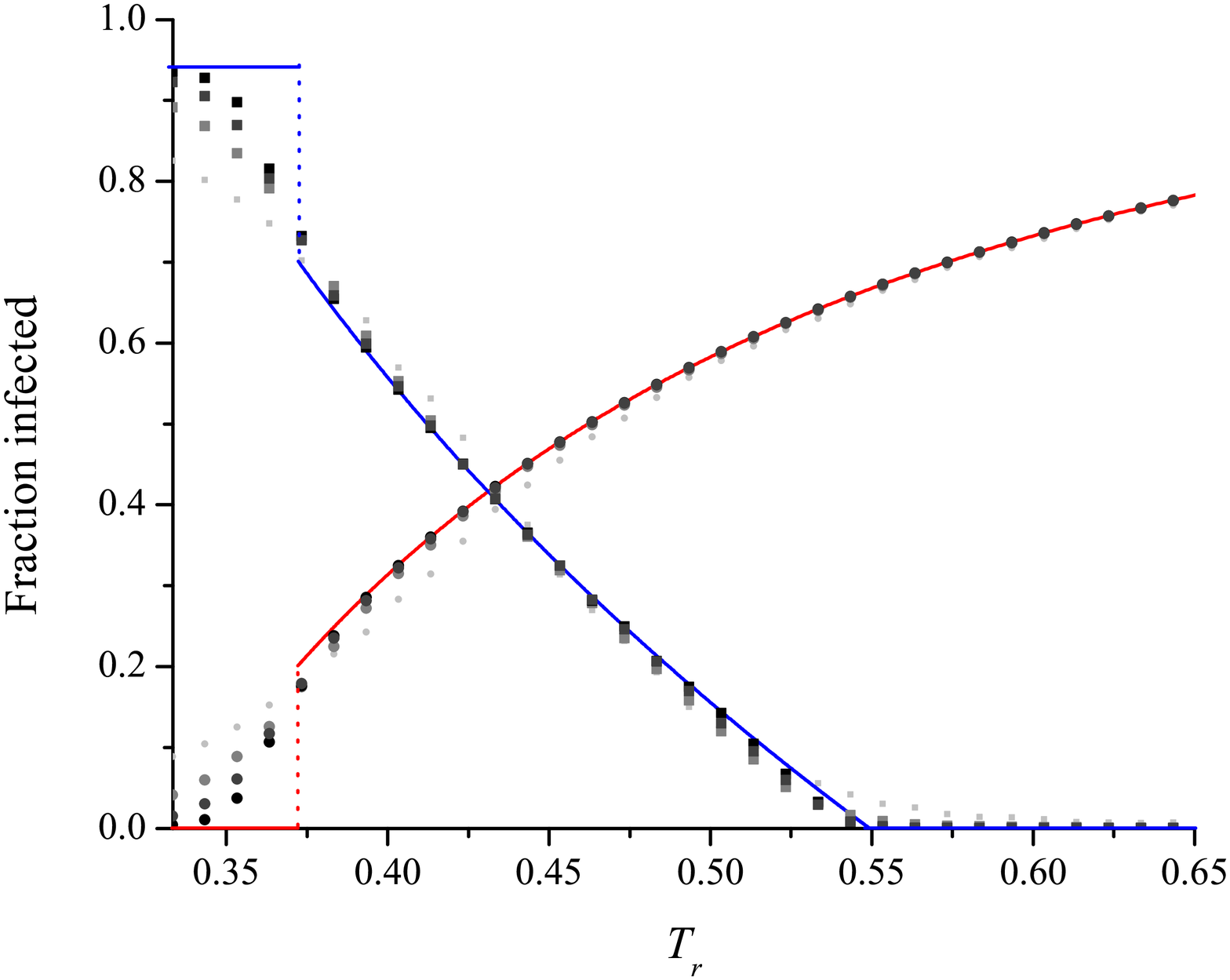}\\
\includegraphics[width=8cm]{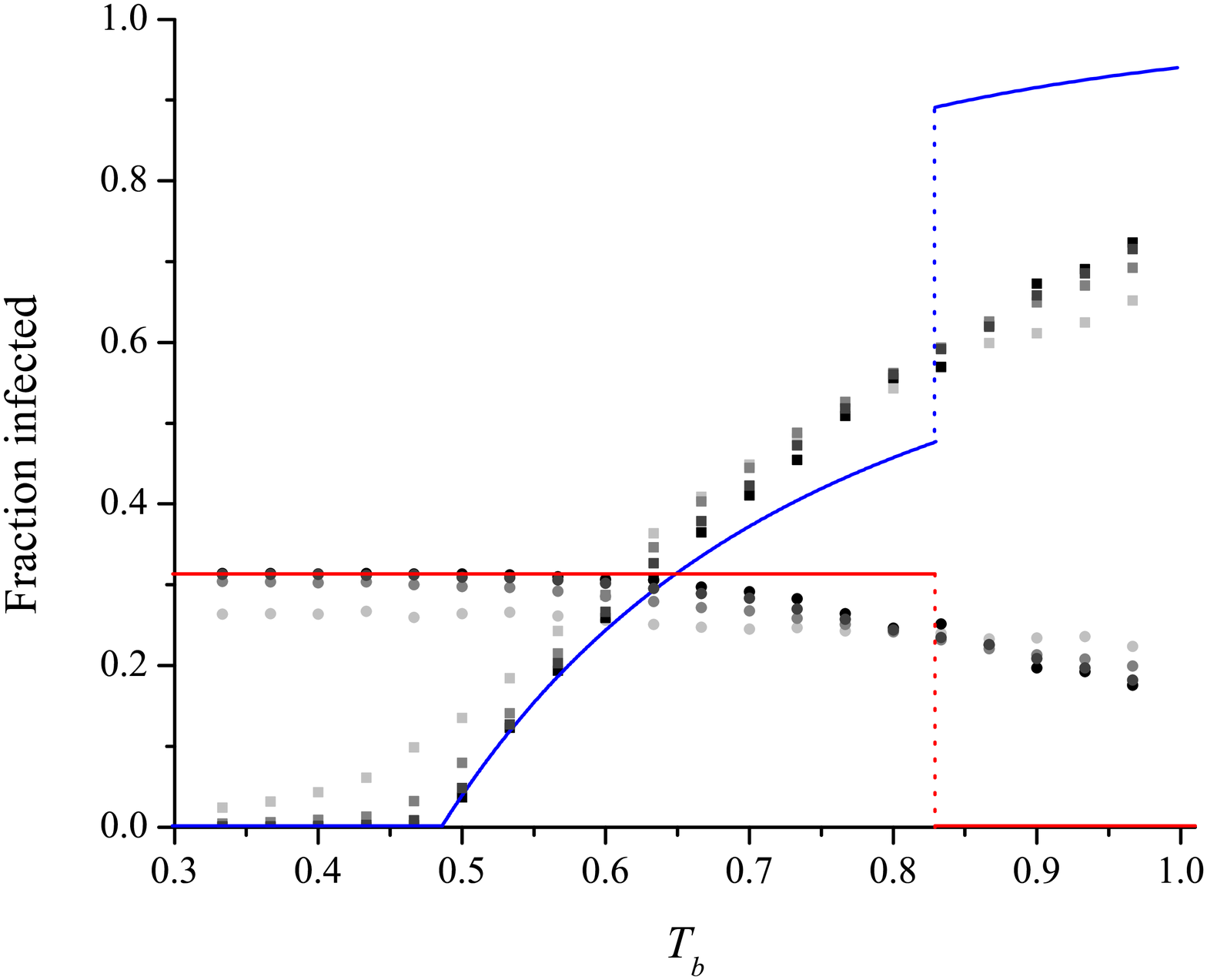}
\end{center}
\caption{Top: simulation results for the average fraction of vertices
  infected with red (circles) and blue (squares) as a function of~$T_r$ for
  the case $\alpha=0.1$, $T_b=1$.  Bottom: the corresponding plot as a
  function of $T_b$ for $T_r=0.4$, $\alpha=0.2$.  The solid lines represent
  the analytic predictions.  The parameter ranges in each case were chosen
  to overlap the coexistence region so that two transitions are visible in
  each panel, the discontinuous transition at the growth-rate boundary and
  the continuous transition at the coexistence threshold.  In the central
  region of each panel the two diseases coexist.  As in
  Fig.~\ref{fig:abovecoexist}, the shades of the data points indicate
  network sizes of $n=10^3$, $10^4$, $10^5$, and~$10^6$ (lightest points to
  darkest).  Each point is an average over at least $100$ networks.}
\label{fig:scans}
\end{figure}

In Fig.~\ref{fig:scans} we show two further plots of epidemic size, this
time with $\alpha$ fixed and either~$T_r$ (upper panel) or~$T_b$ (lower
panel) varying.  Both plots include parameter ranges that fall within the
coexistence regime and in each we can see both the discontinuous
growth-rate transition and the continuous coexistence transition.  As
expected we see some finite-size rounding of the coexistence transition,
particularly for smaller system sizes, and considerably more dramatic
deviations from the analytic predictions around the growth-rate boundary,
as in Fig.~\ref{fig:abovecoexist}.

\section{Epidemic sizes close to the growth-rate boundary}
\label{sec:finitesize}
The numerical results of Section~\ref{sec:simulation} agree well with our
analytic predictions, except in the region close to the growth-rate
boundary, where, as we have seen, the combination of finite-size effects
and fluctuations in the early stages of the growth process can produce
significant deviations from the expected behavior.  In this regime the
large-$n$ theory breaks down.  We can, however, still derive some useful
analytic results.  As we show in this section, even though it is not
possible to predict the final size of either of the two epidemics close to
the growth-rate boundary, the sizes are still related to one another, one
being large whenever the other is small, and we can derive constraints on
the particular combinations of sizes allowed.

Again our calculations are for the configuration model.  We consider a
vertex anywhere in the network and one of the edges connected to that
vertex, and we define~$w$ to be the average probability that neither of the
two diseases was transmitted to the vertex down that edge.  By transmission
we here mean that the pathogen was spread, but not that the vertex
necessarily became infected---a vertex that was previously infected with
either disease will not become infected again even if the pathogen is
spread to it.

The probability that the red disease was spread over the edge in question
is equal to the probability, denoted~$q_r$, that the vertex at the other
end was infected with red, times the probability that transmission
occurred, which is~$T_r$ by definition, for a total probability of
$T_rq_r$.  Similarly the probability for blue is~$T_bq_b$, and the total
probability $1-w$ that either disease is transmitted is~$T_rq_r+T_bq_b$, so
\begin{equation}
w = 1 - T_rq_r - T_bq_b.
\label{eq:defsw}
\end{equation}

Now suppose that the vertex at the end of the edge has excess degree~$k$.
Then the probability that it is infected with neither red nor blue is equal
to the probability that neither disease was transmitted to it along any of
its $k$ edges, which is~$w^k$.  Averaging over the excess degree
distribution~$q_k$, the vertex's average probability of being uninfected is
thus $\sum_k q_k w^k = F_1(w)$, where $F_1$ is the generating function for
the excess degree distribution, as previously.  Then the probability that
the vertex \emph{is} infected is $1-F_1(w)$, which is necessarily equal to
$q_r+q_b$.  Thus we have
\begin{equation}
F_1(w) = 1 - q_r - q_b.
\label{eq:f1w}
\end{equation}
Substituting for $w$ from Eq.~\eqref{eq:defsw}, we then have
\begin{equation}
F_1(1-T_rq_r-T_bq_b) = 1 - q_r - q_b.
\label{eq:surface}
\end{equation}
This equation does not uniquely determine either of the probabilities~$q_r$
or~$q_b$, and indeed, as we have seen, for finite-size networks they are
typically not determined but can take a range of values depending on chance
fluctuations.  But Eq.~\eqref{eq:surface} gives us a relation between the
two such that if we know either one then we know the other.

\begin{figure*}
\begin{center}
\hfill
\includegraphics[width=5cm]{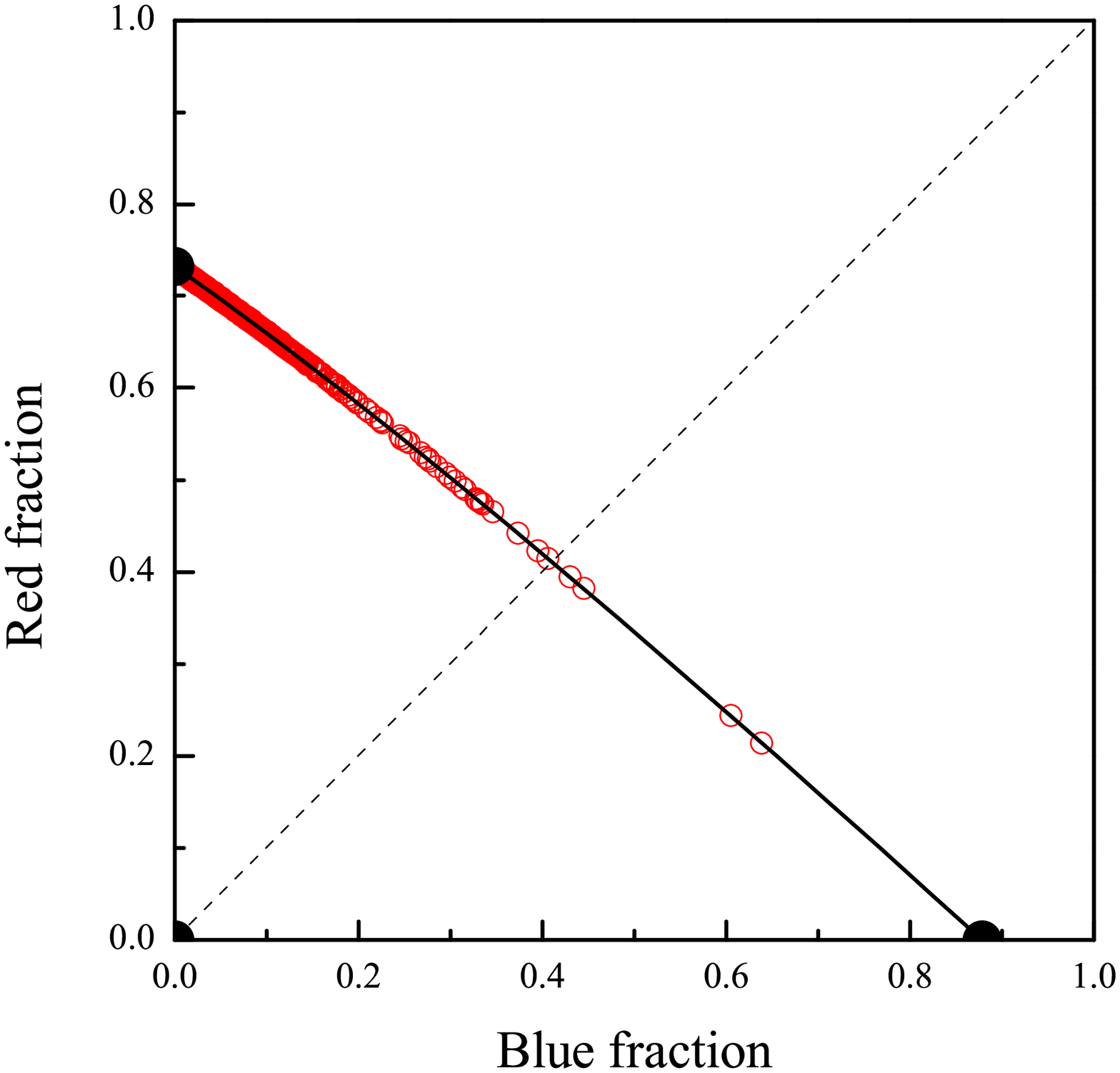}\hfill
\includegraphics[width=5.74cm]{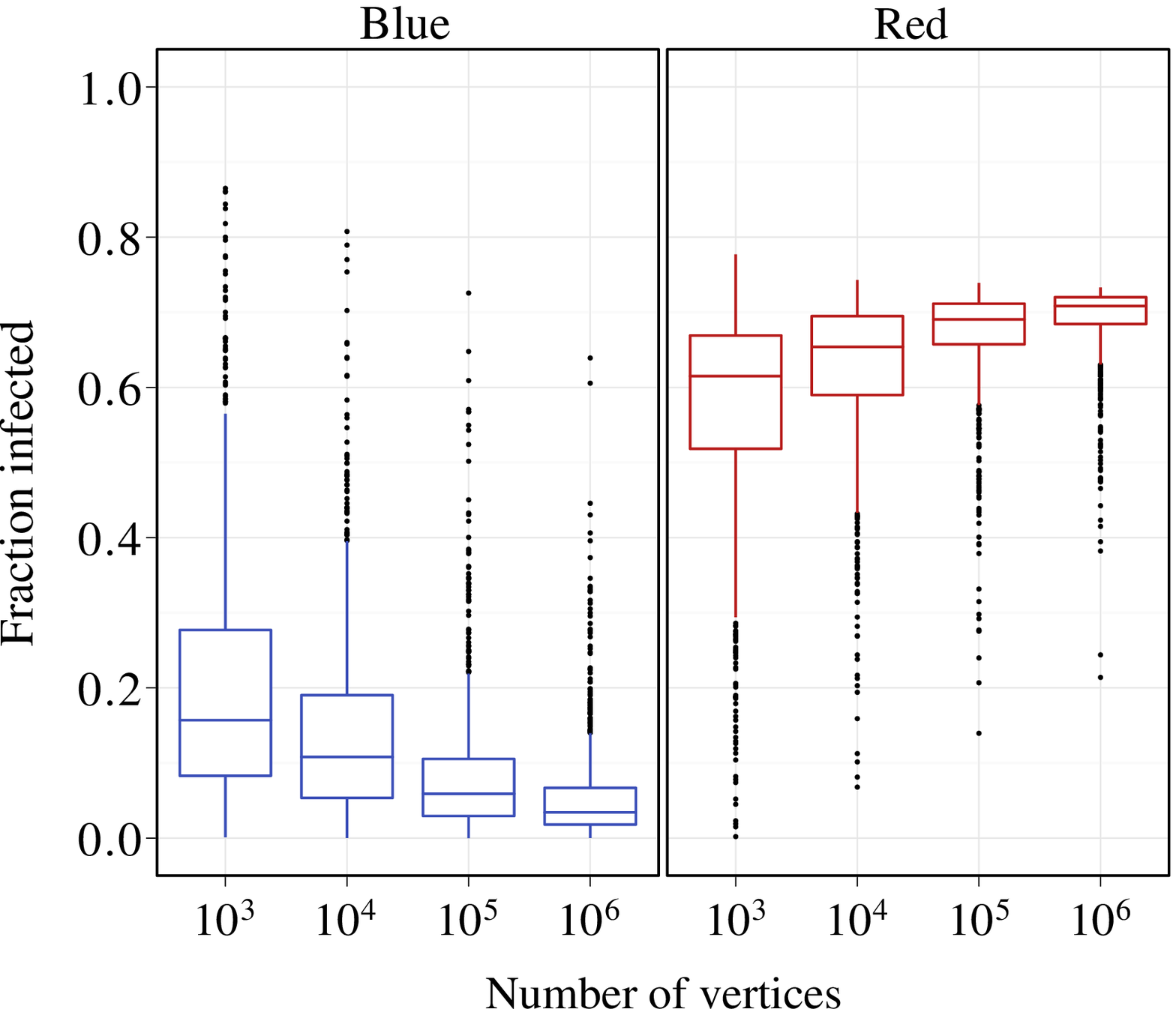}\hfill
\includegraphics[width=5cm]{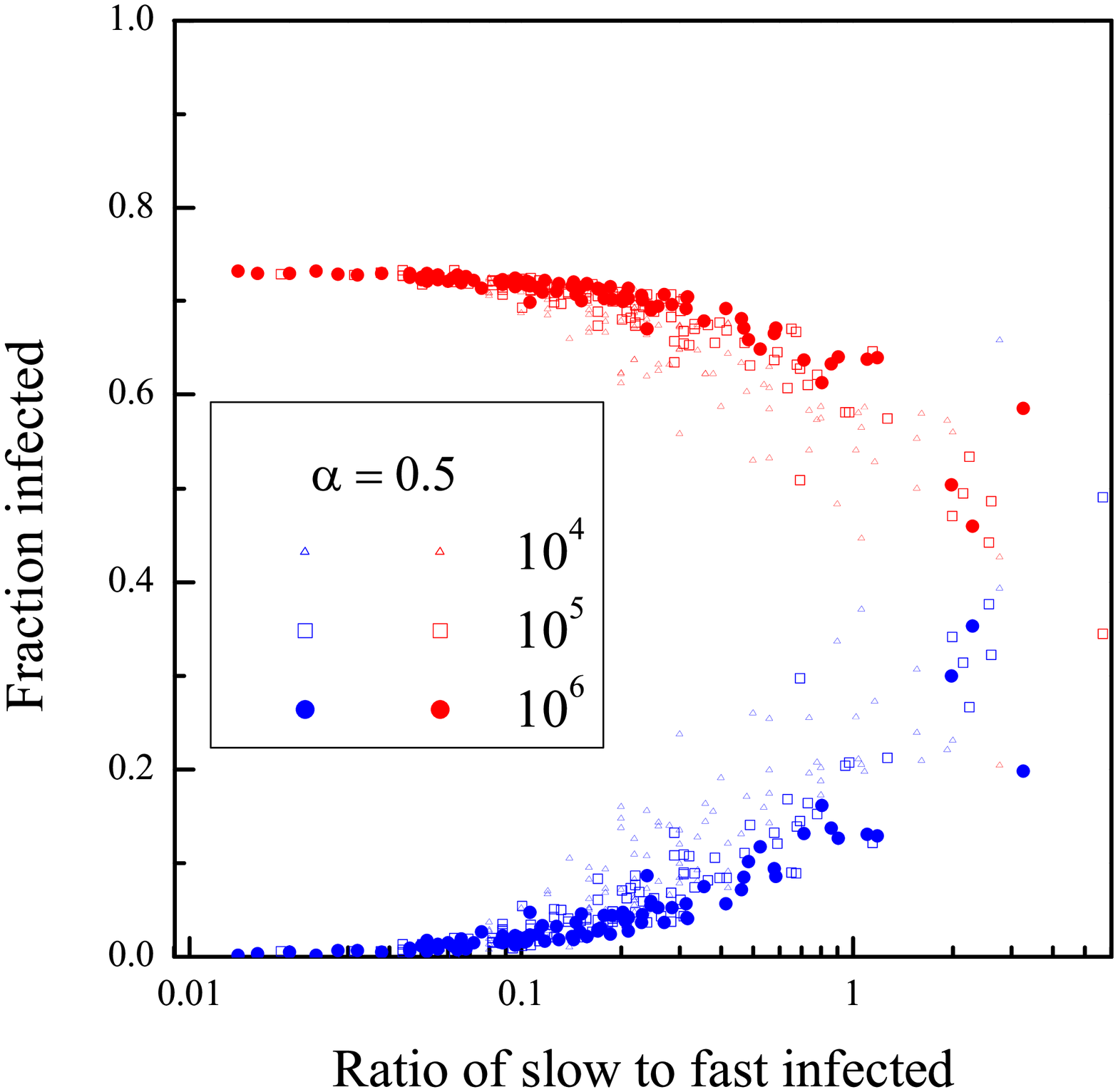}
\hfill\null\\
\hfill
\includegraphics[width=5cm]{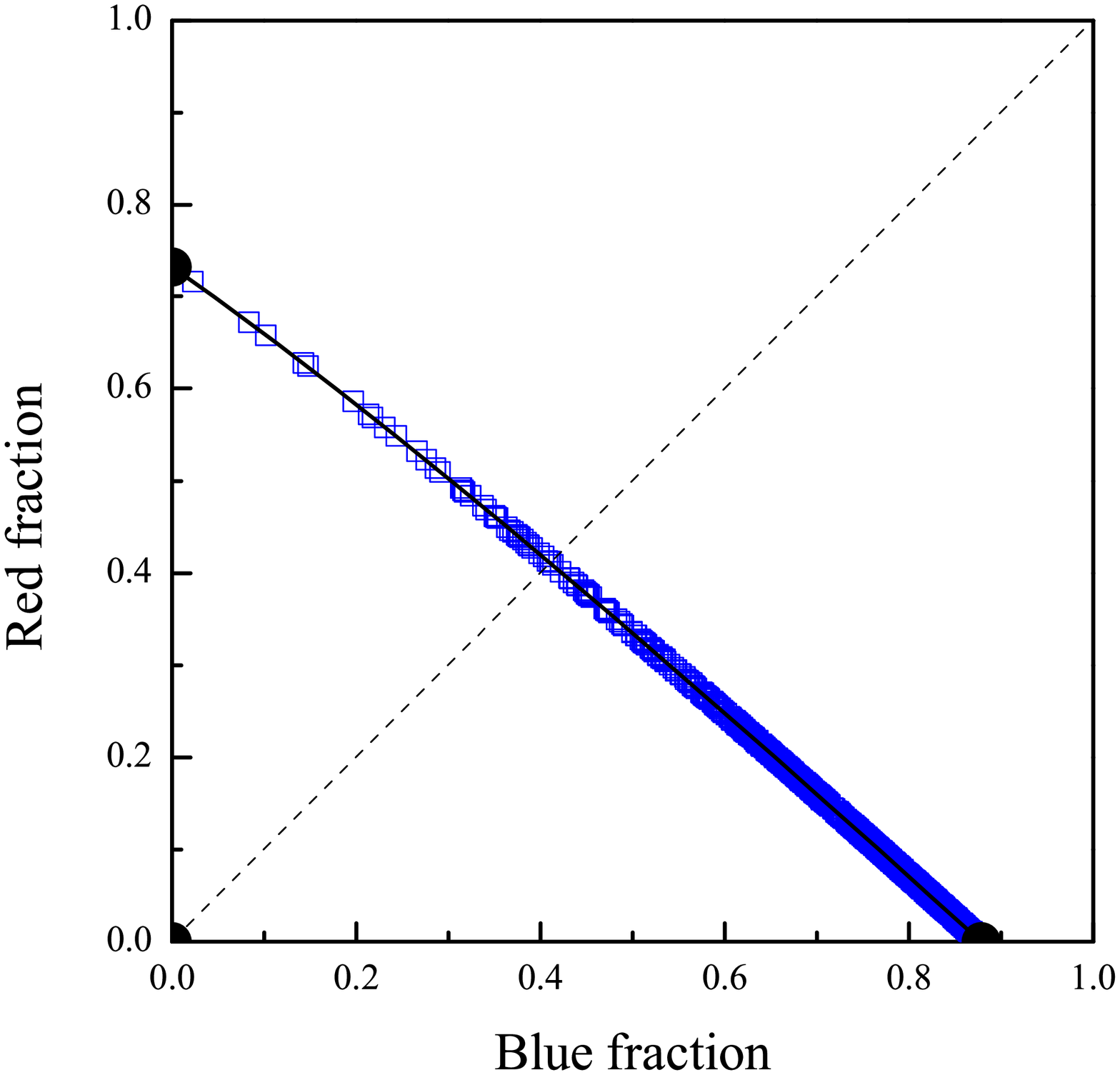}\hfill
\includegraphics[width=5.74cm]{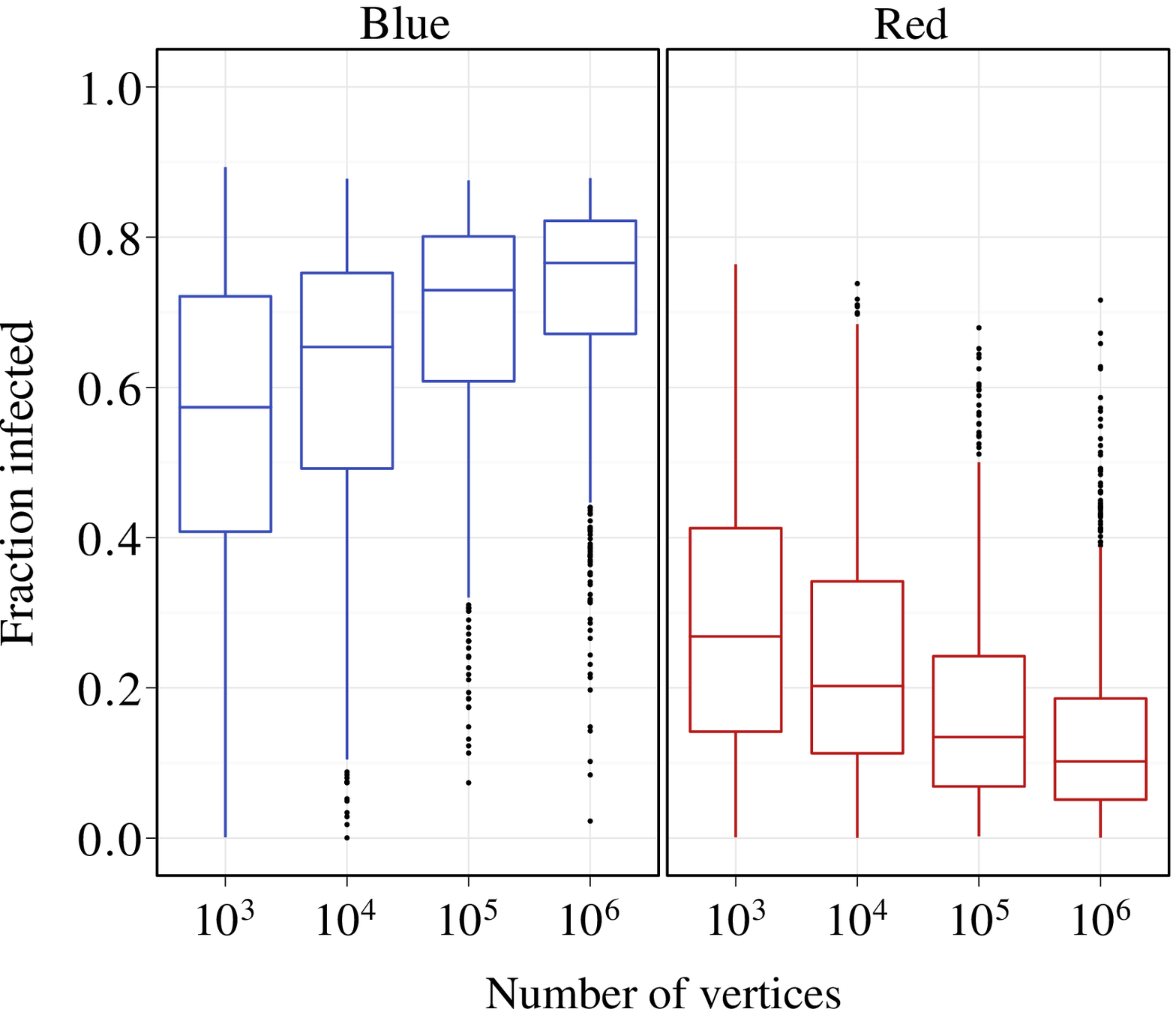}\hfill
\includegraphics[width=5cm]{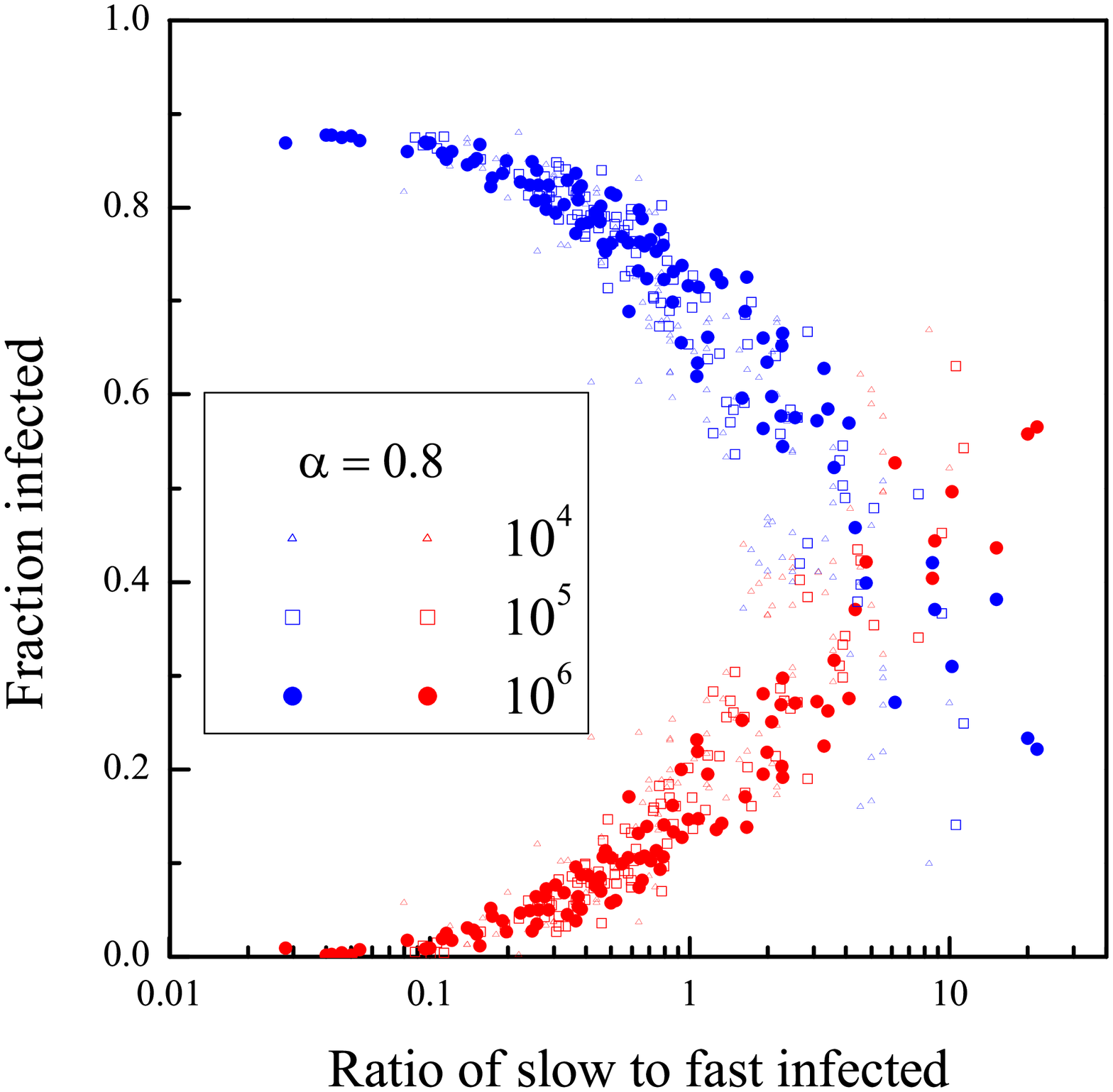}
\hfill\null\\
\hfill
\includegraphics[width=5cm]{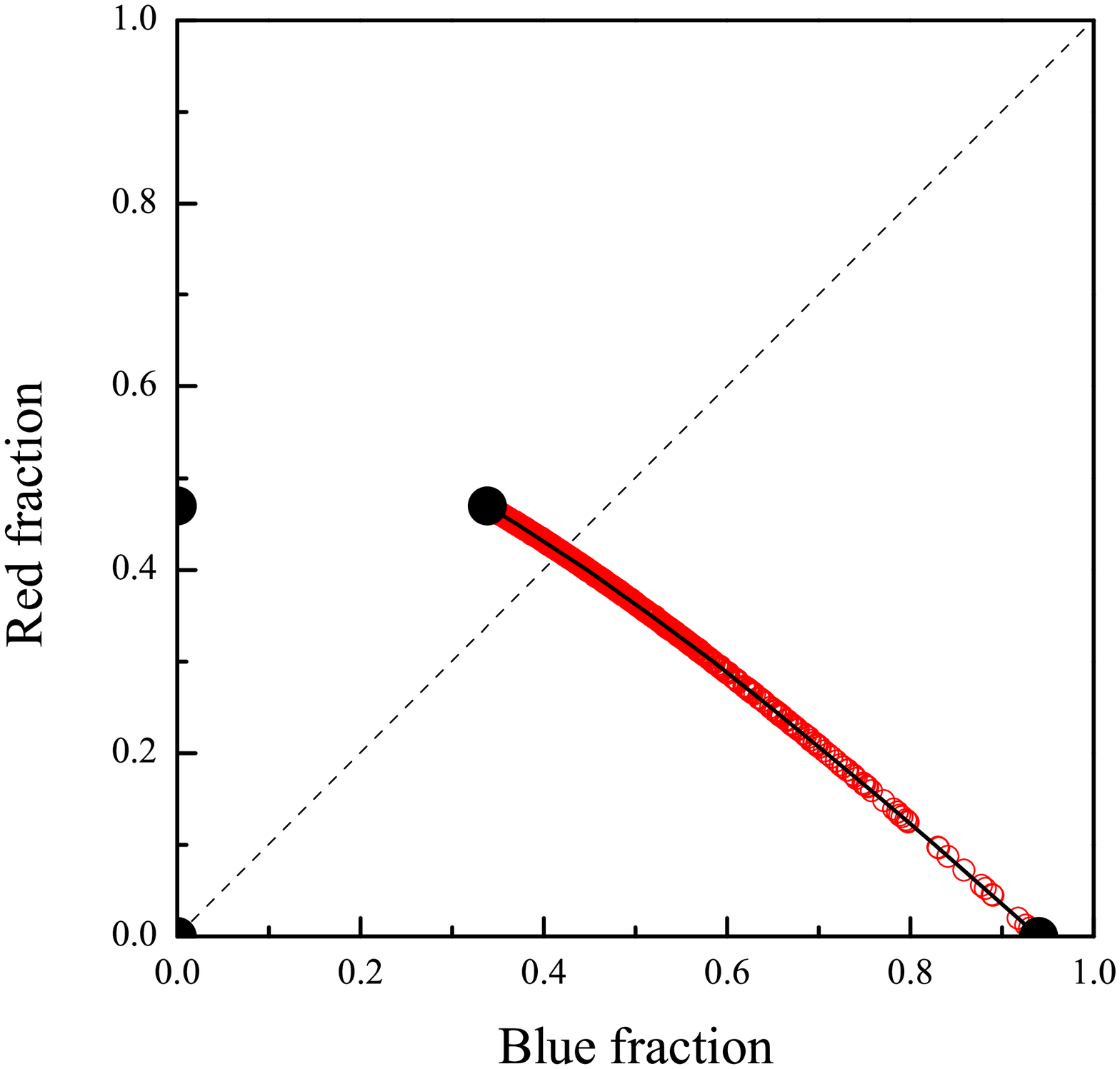}\hfill
\includegraphics[width=5.74cm]{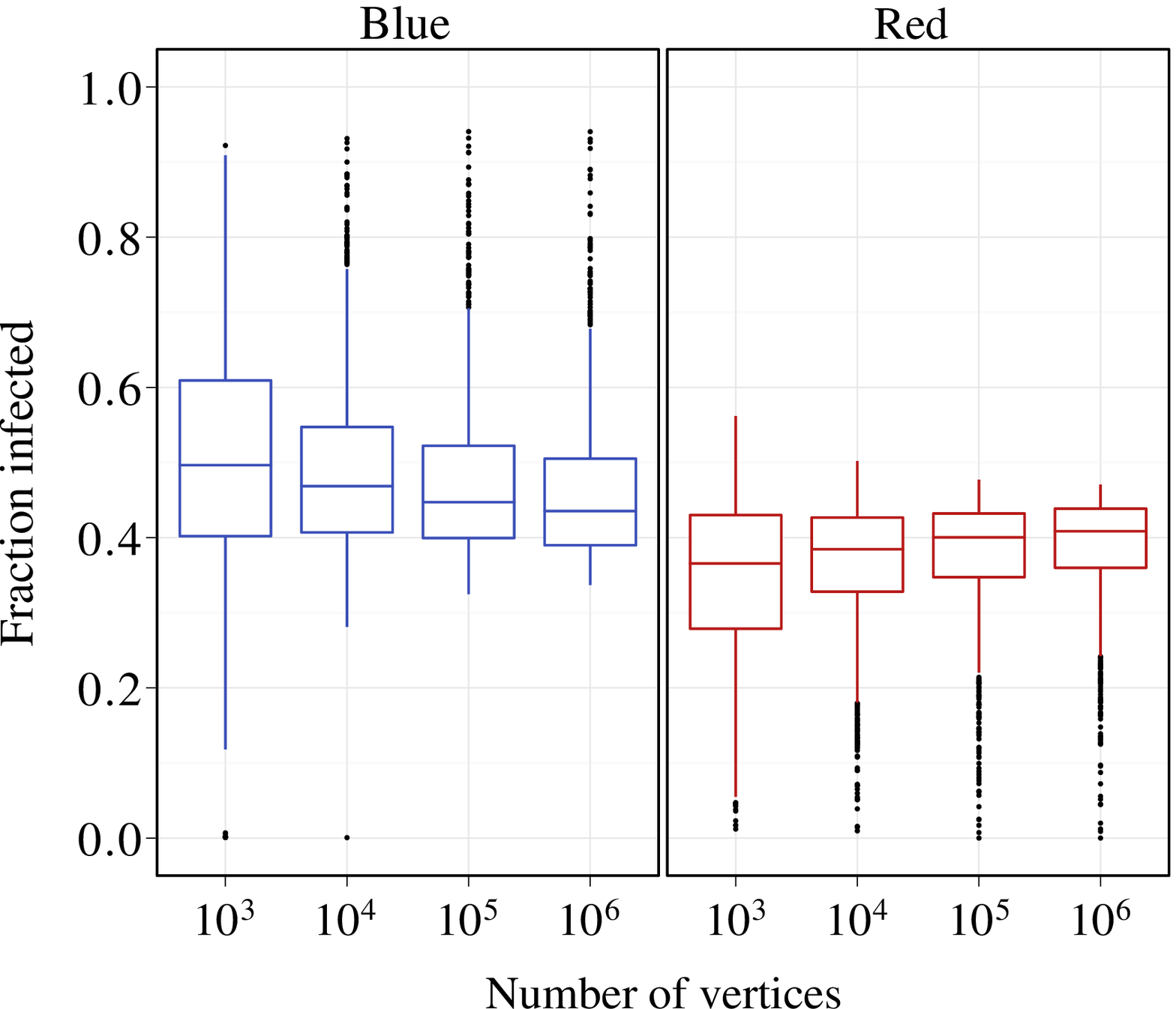}\hfill
\includegraphics[width=5cm]{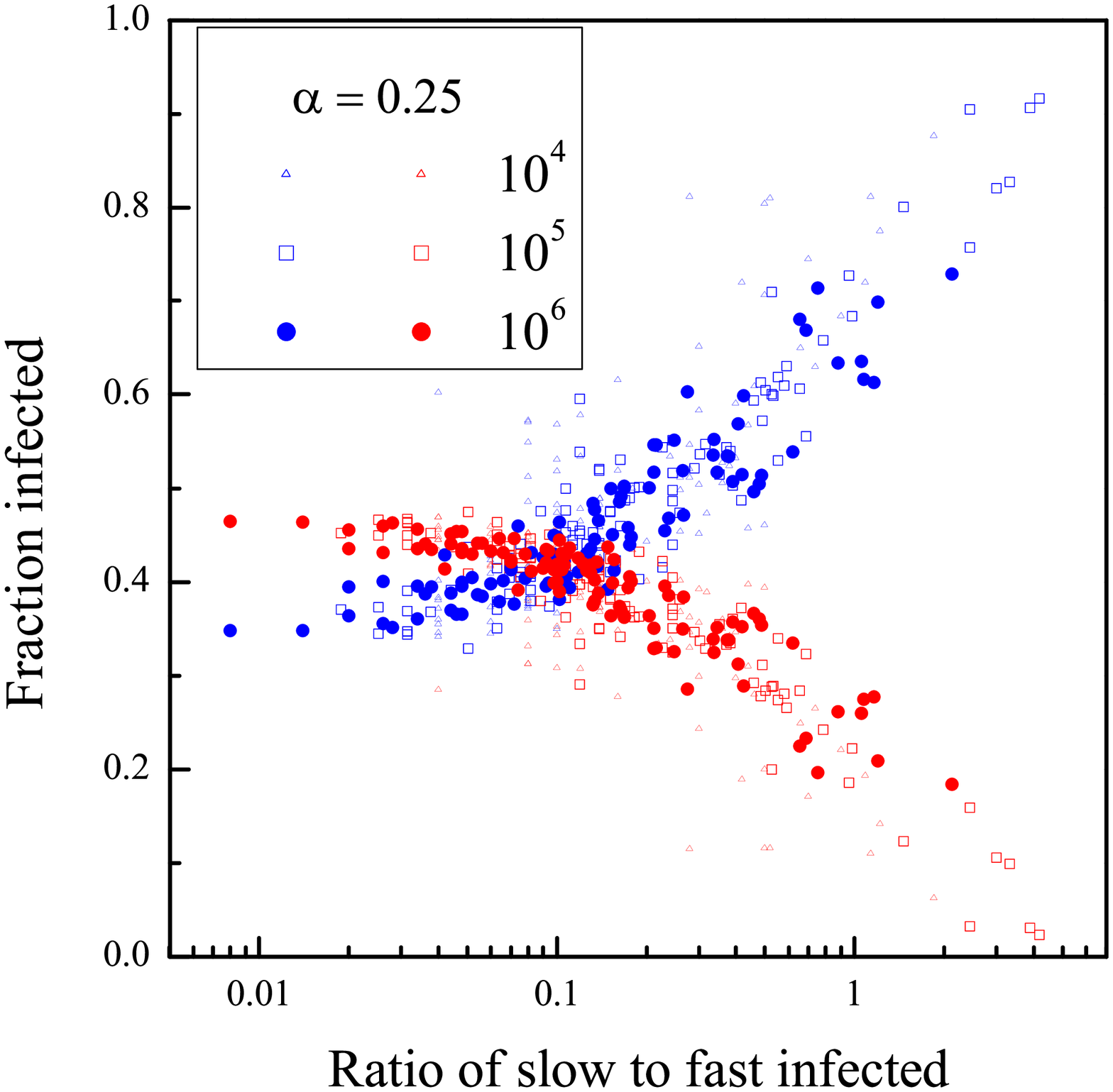}
\hfill\null\\
\hfill
\includegraphics[width=5cm]{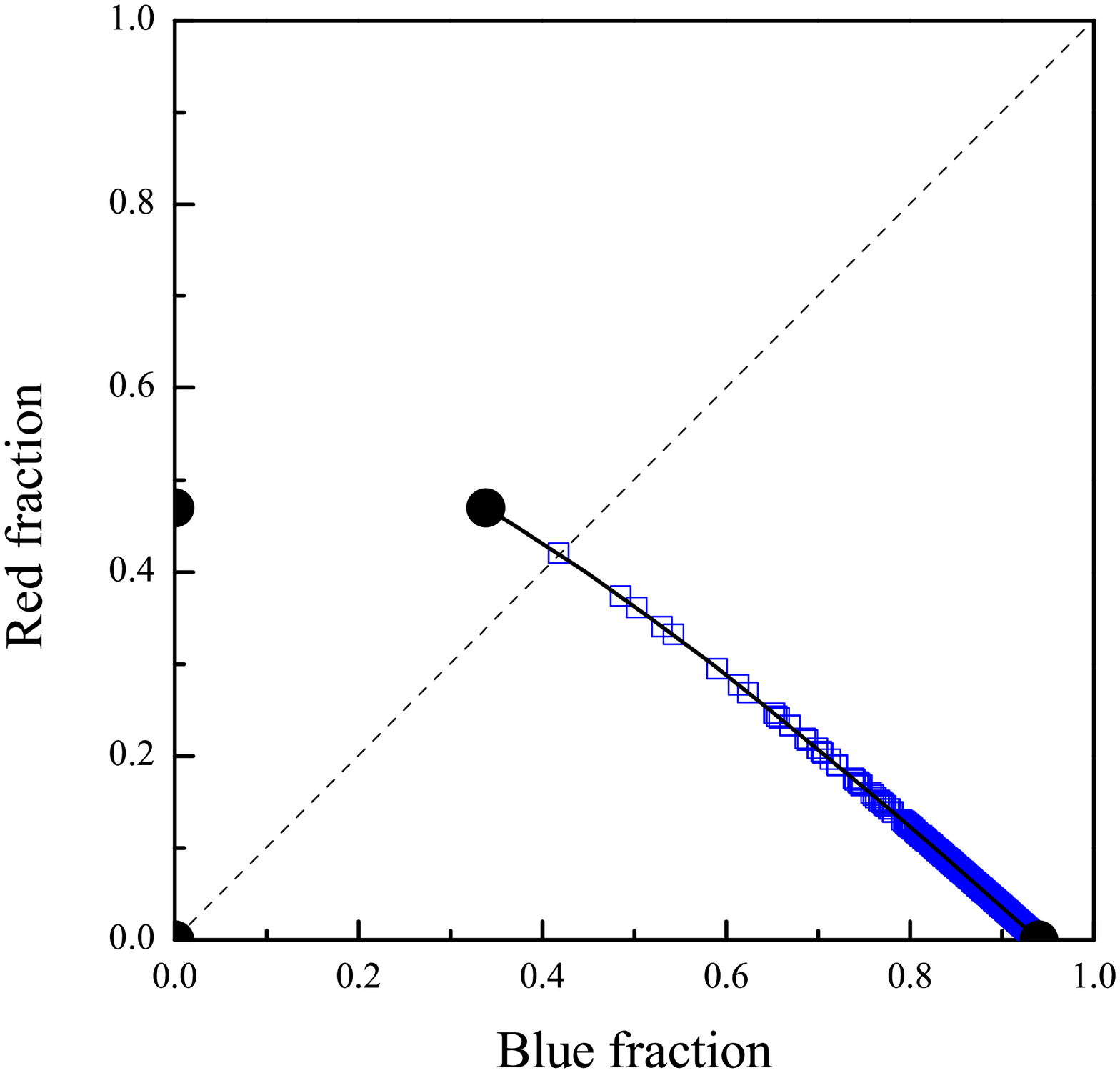}\hfill
\includegraphics[width=5.74cm]{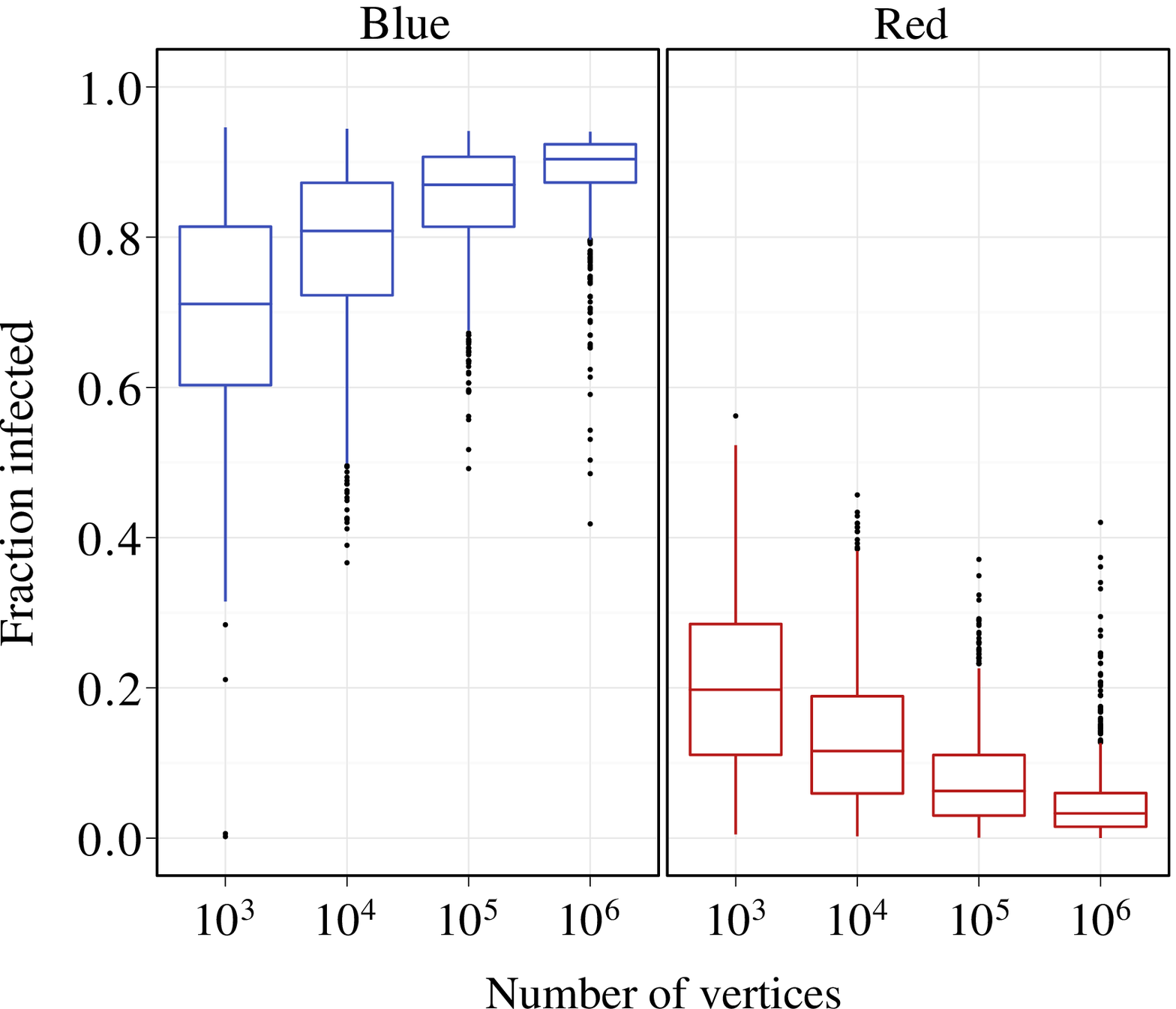}\hfill
\includegraphics[width=5cm]{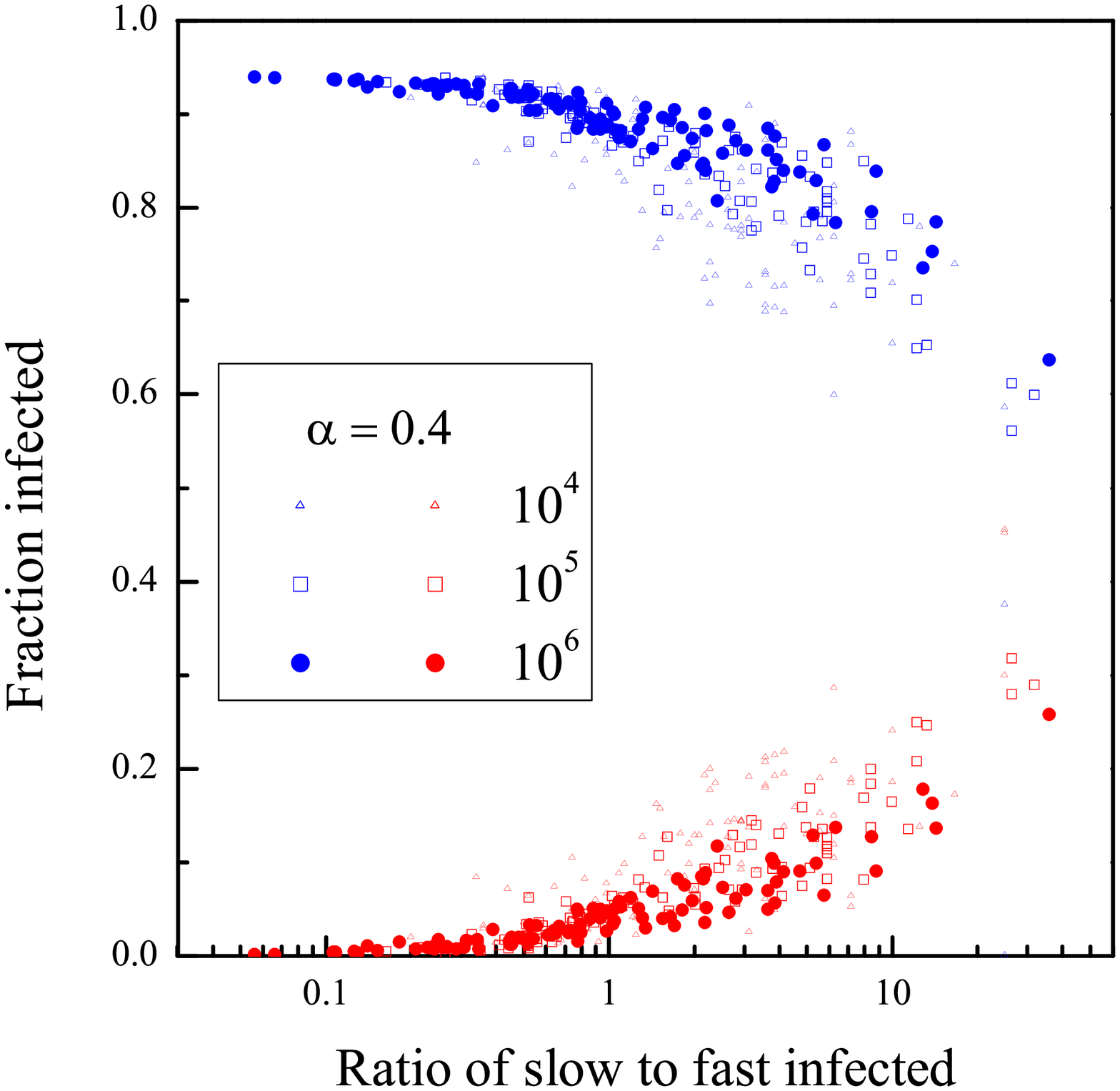}
\hfill\null\\
\end{center}
\caption{Simulation results comparing the fractions of vertices infected
  with the red and blue diseases for four different sets of parameter
  values close to the growth-rate boundary (the four rows in the figure).
  Left column: the fraction infected with red versus blue for networks with
  $n=10^6$.  All points lie on the curves defined by
  Eq.~\eqref{eq:surface}, which are shown as the solid black lines.  Middle
  column: box plot of the fraction infected with each disease for various
  network sizes~$n$.  Right column: the fraction infected with each disease
  at the end of the simulation as a function of the ratio of the numbers
  infected by the faster- and slower-growing diseases measured at the first
  time that either disease reaches $\half\sqrt{n}$.  First row: $T_b =
  0.8$, $T_r = 0.6$, $\alpha = 0.5$; second row: $T_b = 0.8$, $T_r = 0.6$,
  $\alpha = 0.8$; third row: $T_b = 1$, $T_r = 0.45$, $\alpha = 0.25$;
  fourth row: $T_b = 1$, $T_r = 0.45$, $\alpha = 0.4$.}
\label{fig:runs}
\end{figure*}

In Fig.~\ref{fig:runs} we test Eq.~\eqref{eq:surface} against the results
of numerical simulations.  We again take a Poisson degree distribution, for
which, conveniently, since the degree and excess degree distributions are
identical, the probabilities~$q_r$ and~$q_b$ that the vertex at the end of
an edge is infected with red or blue are equal to the overall fractions
$p_r$ and $p_b$ of red and blue vertices in the network as a whole.  For
four different sets of values of the model parameters close to the
growth-rate boundary (the four rows in Fig.~\ref{fig:runs}) we measure
these two fractions on 1000 different networks and in the left column of
Fig.~\ref{fig:runs} we show scatter plots of $p_r$ against~$p_b$.  The
solid lines in the plots represent Eq.~\eqref{eq:surface} and, as we can
see, the values of $p_r$ and $p_b$ indeed lie along this line.  On any
particular run of the simulation one cannot predict where the individual
values $p_r$ and $p_b$ will fall, but if we know one then we can predict
the other, since they always fall on this line.  (The lines appear straight
in the figure, but are in fact slightly curved for this particular choice
of degree distribution and model parameters.)

Of the four rows in the figure, the first two have the same values of $T_r$
and $T_b$, chosen to fall above the coexistence threshold in the regime
where coexistence does not occur, and two different values of $\alpha$
chosen to fall a little below and a little above the growth-rate boundary.
Thus for the parameter values in the first row the red disease is faster
growing and would dominate in the limit of large~$n$ but the blue one
occasionally wins on the finite network.  The second column gives a box
plot showing the average fraction infected with each disease for a range of
system sizes~$n$, and we can see that as $n$ becomes larger the dominance
of red becomes progressively better defined.  In the second row of the
figure blue is the faster growing disease and the positions are reversed,
blue winning more often, particularly as $n$ becomes large.

The third and fourth rows of Fig.~\ref{fig:runs} show similar results, but
for $T_r$ below the coexistence threshold.  Now when $\alpha$ is chosen to
put us below the growth-rate boundary, in the coexistence region (third row
in the figure), we see that neither red nor blue dominates, even for
large~$n$, with both infecting significant fractions of the network.  In
the fourth and final row of the figure the value of $\alpha$ places us
above the growth-rate boundary again and in this regime blue dominates in
the limit of large~$n$.

These results accord well with the analysis presented earlier.  We have
not, however, directly tested our hypothesis that the deviations we observe
from the infinite-$n$ results are due to stochastic fluctuations in the
early stages of the growth process.  The third column of
Fig.~\ref{fig:runs} shows a test of this hypothesis.  The test involves
waiting an initial period of time for the fluctuations to become small,
then measuring the number of individuals infected with each disease.  The
horizontal axis in each panel of the third column gives the ratio of the
number of vertices infected by the disease (either red or blue) with the
slower exponential growth rate versus the number infected by the disease
with the faster, measured at the first moment that either disease reaches
$\half\sqrt{n}$ infected individuals.  The vertical axis measures the final
fraction of individuals infected with each disease, once both have run
their course, and the scatter of red and blue points shows the results for
each of our simulations, for a range of different network sizes.

On the left-hand side of each graph in the third column the ratio of slow-
to fast-growing diseases is small, meaning that the fast-growing one
dominates at early times.  In these circumstances, if we assume that
fluctuations are no longer important to the fast-growing disease and that
it simply grows exponentially with the appropriate growth rate, then the
final size of its outbreak is uniquely determined, and hence so is the
final size of the outbreak of the slower disease.  The results in the
figure appear to confirm this conjecture, with the observed fractions
infected with each disease being narrowly concentrated around single
values.  As we move further to the right, we enter the regime in which the
slower growing disease infects more individuals at early times.  In this
regime, and particularly in the rightmost regions of the graph, the faster
growing disease only infects a small number of individuals and hence can
have significant fluctuations.  These fluctuations may give the
fast-growing disease a boost sufficient to allow it to catch up to the
slower one, or a delay sufficient to ensure that it does not.  Thus in this
region the outcome is not uniquely determined and we expect to see a range
of final infected fractions.  Again the numerical results shown in the
figure appear to confirm this conjecture, with a much broader scatter of
points on the right than on the left.  Finally, note that the points in the
right-hand part of the figure are primarily those for simulations on
smaller networks (as indicated by the shapes of the points), because larger
systems take longer to reach the $\half\sqrt{n}$ point, making the typical
values of the slow-to-fast ratio (which dwindles over time) smaller and
hence pushing the points leftward on the plot as $n$ grows.

\section{Conclusion}
In this paper we have studied the behavior of two competing diseases with
complete cross-immunity (or two strains of a single disease) spreading
concurrently over a static network of contacts between individuals of a
single population.  Using a mixture of analytic results, heuristic
arguments, and numerical simulation, we have derived the phase diagram for
the system, which shows four distinct phases, and given calculations of the
expected number of individuals infected with each of the diseases in the
limit of large network size for networks generated using the configuration
model.  Of particular interest is the coexistence phase, a region of the
parameter space in which neither disease excludes the other and both spread
to infect an extensive fraction of the network.  We have demonstrated a
number of nontrivial features of this phase, including the fact that the
disease that expands through the population slower must nonetheless have a
higher probability of transmission if coexistence is to occur.  Such
behavior is possible only if the faster growing disease has a shorter
transmission time between when an individual gets infected and when they
pass on the infection to others, and we have shown also that there is an
upper bound on the ratio of the transmission times of the two diseases if
coexistence is to occur.

Another unusual feature of the system is the ``growth-rate boundary,'' a
dynamical transition between regimes in which one disease or the other
dominates.  The numbers of individuals infected by both diseases change
discontinuously as we cross this boundary, in the limit of large system
size, although for finite systems the transition is blurred by strong
finite-size effects.

A number of extensions or generalizations of our calculations are possible.
We have worked with the simplest possible model of epidemic dynamics, the
Reed--Frost model, but calculations could be performed for other SIR-style
models, such as the Kermack--McKendrick model in which infection and
recovery processes take place in continuous time with stochastically
constant rates.  The results should be qualitatively similar: one expects
to see a growth-rate boundary in any model with exponential growth rates.
Extending the model further, one could introduce arbitrary distributions of
transmission and recovery times~\cite{KN10a}, but determining the
exponential growth rates will be more complicated for such a system.
Changes of this kind may alter the shape of the growth-rate boundary but
should not change the qualitative nature of the transition.

There are also some questions of interest concerning the Reed--Frost model
that are not answered by the calculations presented here.  Our analytic
results are all derived for the configuration model and different behaviors
might be seen on other networks.  Also, close to the growth-rate boundary
we observe strong finite-size deviations from the large-$n$ predictions of
the analytic theory, and it is unclear at present how to calculate, for
instance, the exponents characterizing the growth of epidemics with system
size in this regime.  And we have assumed in our calculations that neither
disease dies out in the early stages of the growth process; in reality they
will sometimes die out and we do not at present know how to calculate the
probabilities of certain outcomes, such as the probability within the
coexistence region that the slower-growing disease will die out after the
faster-growing one has spread.  This probability depends on the chance that
the slower disease reaches the giant component of the residual network, a
probability that does not appear to have a simple expression in the
percolation-theory language employed here.  These and other interesting
questions we leave for future work.

\begin{acknowledgments}
  This work was funded in part by the National Science Foundation under
  grant DMS--0804778 and by the James S. McDonnell Foundation.
\end{acknowledgments}


\begin{thebibliography}{10}
\bibitem{note1}
  For technical reasons we also require that the first and second moments
  of the degree distribution be finite.
\bibitem{note2}
  The definition of the coexistence threshold used here differs slightly
  from that used in Ref.~\cite{Newman05c}.  In that paper the threshold was
  defined as the transmissibility~$T_r$ above which coexistence is
  impossible for \emph{any} value of~$T_b$, no matter how large, whereas it
  is defined here to be the transmissibility for the \emph{actual} value
  of~$T_b$.  Equivalently, the definition used in~\cite{Newman05c} is equal
  to our definition when $T_b$ takes its maximum value of~1.
\bibitem{note3}
  Strictly speaking, at $t=1$, Eqs.~\eqref{eq:nbtime} and~\eqref{eq:nrtime}
  should read $N_b = T_b \av{k}$ and $N_r = T_r \av{k}$ because the initial
  carriers of the diseases are chosen at random and not infected via a
  neighbor.  In real life, on the other hand, all infections are acquired
  via a neighbor, so the equations should always be correct.  These
  subtleties do not affect our arguments, however, so we ignore them here.
\bibitem{note4}
  A naive implementation of this scheme would be computationally wasteful,
  and also somewhat arbitrary since one would have to decide a threshold
  level of infection below which a disease would be deemed to have died
  out.  In our simulations we mitigate these problems somewhat by first
  generating the bond percolation clusters for the entire network and
  starting the simulated spread of the two diseases at vertices chosen from
  the giant clusters.  This does not guarantee that both diseases will
  produce an epidemic in the coexistence region but it reduces
  substantially the number of cases where a disease dies out.
\end{thebibliography}
\end{document}